\documentclass[english,12pt,aps,prd,a4paper,preprintnumbers,floatfix,nofootinbib,superscriptaddress, notitlepage]{revtex4} 
\usepackage{graphicx}
\usepackage[utf8]{inputenc} 
\usepackage{amsmath}
\usepackage{amssymb}
\usepackage{float}
\usepackage{comment}
\usepackage{slashed}
\usepackage[normalem]{ulem}
\usepackage{braket}
\usepackage{caption}
\usepackage{adjustbox}
\usepackage[usenames,dvipsnames]{color}
\usepackage[colorinlistoftodos]{todonotes}
\usepackage[colorlinks=true,citecolor=darkred,urlcolor=darkred, pdfborder={0 0 0}]{hyperref}
\definecolor{darkred}{rgb}{0.6,0,0}
 
\newcommand {\ignore}[1]{}

\newcommand{\bea}{\begin{eqnarray}}
\newcommand{\eea}{\end{eqnarray}}

\def\gsim{\raise0.3ex\hbox{$\;>$\kern-0.75em\raise-1.1ex\hbox{$\sim\;$}}}
\def\lsim{\raise0.3ex\hbox{$\;<$\kern-0.75em\raise-1.1ex\hbox{$\sim\;$}}}

\usepackage{soul}

\definecolor{mightnightblue}{RGB}{25,25,112}

\definecolor{brown}{rgb}{0.59, 0.29, 0.0}

\def\21{$\mathrm{SU(2)_L \otimes U(1)_Y}$}

\usepackage{mathrsfs}

\newcommand{\AddrIISERB}{Department of Physics, Indian Institute of Science Education and Research - Bhopal, \\ 
Bhopal Bypass Road, Bhauri, Bhopal 462066, India}
\begin{document}

\title{\textcolor{BrickRed}{$B-L$ model in light of the CDF II result}}
\author{Sanjoy Mandal}\email{smandal@kias.re.kr}
\affiliation{Korea Institute for Advanced Study, Seoul 02455, Korea}
\author{Hemant Prajapati}\email{hemant19@iiserb.ac.in}
\thanks{Author to whom any correspondence should be addressed.}
\affiliation{\AddrIISERB}
\author{Rahul Srivastava}\email{rahul@iiserb.ac.in}
\affiliation{\AddrIISERB}
\date{\today}
\begin{abstract}
Recent CDF II collaboration's result on $W$ mass measurements contradict Standard Model prediction, requiring new physics to explain this anomaly. Such new physics may manifest through tree-level or loop-level corrections to the mass of the $W$ boson. In this work, we investigate the possibility that the CDF-II result is indicative of new physics not directly changing the $W$ boson mass but rather the $Z$ boson mass. Since the $Z$ boson mass goes as an input into the Standard Model prediction for $W$ boson mass, this change in $Z$ mass ultimately leads to the discrepancy between the CDF-II measurement and the Standard Model expectation.    We demonstrate this idea through one of the simplest and most studied $U(1)$ gauge extensions of the Standard Model, namely the gauged $U(1)_{B-L}$ extension. We demonstrate that $B-L$ extended models can explain the revised best-fit values for $S$, $T$, and $U$ following the CDF II results. We studied the parameter space of models with and without mixing between neutral gauge bosons. We also reviewed the dark matter constraints and demonstrated that there is parameter space that is compatible with the current $W$ boson mass, relic abundance, and direct detection experiments.
\end{abstract}
\maketitle
\section{Introduction}
\label{sec:intro}
The Standard Model~(SM) of electroweak theory with $SU(2)\otimes U(1)$ gauge symmetry is highly successful in explaining most of the observations in particle physics experiments. The recent finding of the Higgs-like boson with mass 125 GeV at the LHC~\cite{ATLAS:2012yve,CMS:2012qbp},  seems to complete the SM. Despite its ability to explain most observable phenomena, the SM cannot be considered the final theory of the particle physics. There is an ever-increasing number of observations, such as the discovery of neutrino oscillations~\cite{Super-Kamiokande:1998kpq} and
the existence of dark matter~(DM) at cosmic scales~\cite{Planck:2018vyg}, that put serious questions on the SM predictions. In addition to this, the CDF-II collaboration recently published their high precision measurement of the $W$ boson mass $M_W^{\rm CDF} = 80.4335 \pm 0.0094$ GeV~\cite{CDF:2022hxs}, which reveals a $7$-$\sigma$ difference from the SM expectation $M_W^{\rm SM} = 80.354 \pm 0.007$ GeV~\cite{ParticleDataGroup:2020ssz}. This is a significant deviation, and new physics is required to account for it. This leads us to investigate the extension of SM, which can account for the aforementioned problems with SM.

\par The initial line of reasoning for CDF II measurements was to consider this to be a purely $W$ mass anomaly and attempt to increase $W$ boson mass at the tree or loop level. Many works have thus proposed new physics models in which the CDF II excess of $W$ boson mass over SM expectation may be explained by new physics modifying $W$ boson mass at the tree level or at the loop level through the existence of additional $SU(2)$ multiplets \cite{Batra:2022arl,CentellesChulia:2022vpz,Wu:2022uwk,Dcruz:2022rjg,Batra:2022pej,Pfeifer:2022yrs,Zhu:2022scj,Popov:2022ldh,Batra:2022org,Balkin:2022glu,Crivellin:2022fdf,Heo:2022dey,Kanemura:2022ahw,Zhang:2022nnh,Borah:2022obi,Cheng:2022aau,Heeck:2022fvl,Baek:2022agi,Borah:2022zim}. Nonetheless, it is possible to believe that this anomaly originated in a different sector before being translated to the $W$ mass. The expressions of the observables in that sector are changed by some anomaly and these modified parameters go as an input to calculate the $W$ mass, hence changing the $W$ mass \cite{Endo:2022kiw,Blennow:2022yfm}. 
Here, we focus on the second type of models, particularly we focus on the $Z$ boson. The mass of the $Z$ boson can be modified through mixing with a new neutral gauge boson, $Z'$, which may arise from the gauging of an additional $U(1)$ symmetry. Such $Z-Z'$ mixing can occur either via kinetic mixing gauge bosons or through a mass-mixing. We argue that the CDF II $W$ anomaly can be thought of as an anomaly in $Z$ boson mass that can be translated to the $W$ mass through oblique parameters.  Furthermore, the new physics contributions can bring the $Z$ mass to the experimentally observed value, hence transferring the anomaly from the $Z$ boson mass to the $W$ boson mass.

\par In the absence of a right-handed neutrino~(RHN), the neutrinos are massless in the SM. RHNs have been a prevalent feature of many extensions of the SM, such as various seesaw mechanisms~\cite{Schechter:1980gr,Schechter:1981cv,Mohapatra:1986bd,Malinsky:2005bi} to generate neutrino masses. In recent years, a number of models have been proposed that combine neutrino mass generations and the existence of DM into a single framework. Motivated by this, people have studied extensively beyond standard model~(BSM) framework based on the gauged $U(1)_{B-L}$ model~\cite{Okada:2016tci,Das:2022oyx,Das:2017deo,Basso:2011hn,Emam:2007dy,Basso:2010jt,Rodejohann:2015lca}. The most intriguing aspect of this model is that it includes three RHNs to cancel gauge and mixed
gauge-gravity anomalies and generate tiny neutrino masses through the seesaw mechanism. One also has the possibility of explaining the DM in these types of models with an additional scalar field, $\chi_d$, that is a SM singlet but charged under $U(1)_{B-L}$.  An advantage of this scenario is that one does not need to impose any ad hoc $\mathcal{Z}_2$ symmetry to stabilize the DM. Instead, the stability of $\chi_d$ can be guaranteed by appropriately choosing its $B-L$ charge. This type of model predicts the existence of a new neutral gauge boson $Z'$ that can mix with the SM neutral gauge boson $Z$~\cite{Langacker:2008yv}. This mixing changes the Standard expression for the Z boson mass in the Standard Model, thus altering the W boson mass.

\par We show that, despite its simplicity, in addition to neutrino mass generation and DM, $B-L$ model can also explain the recent CDF-II $W$ boson mass measurements. The new boson associated with $U(1)_{B-L}$ symmetry mixes with the SM neutral $Z$ boson to provide $S$, $T$, $U$ corrections that are compatible with current $W$ boson mass measurements.  Specifically, we investigate two distinct scenarios: one with no mass mixing between two neutral bosons and one with mass mixing. We study the difference in parameter space in both cases. We show that the parameter space consistent with the best-fit $S$, $T$, $U$ values following the CDF II results is also consistent with the DM physics constraints in the model we proposed.
\par The paper is organised as follows: in Sec.~\ref{sec:KM}, we briefly discuss the possibility of having kinetic mixing between two field strength tensors corresponding to $U(1)_Y$ and $U(1)_{B-L}$. We investigate in detail whether or not addressing simply the impacts of kinetic mixing at the tree level may resolve the $W$ mass anomaly. In Sec.~\ref{sec:model}, we study two $B-L$ gauged models without taking into account the effects of kinetic mixing. The first model is a minimal $B-L$ extension of the SM with no mass mixing between the SM neutral gauge boson $Z$ and $U(1)_{B-L}$ neutral gauge boson $Z'$. In the second model, we introduce mass mixing between these neutral gauge bosons and also introduce a scalar DM candidate. In Sec.~\ref{sec:wstu}, we discuss how one can parametrise the new physics contributions to $W$ mass in terms of oblique parameters $S$, $T$ and $U$. In Sec.~\ref{sec:STUspace}, we described the effective Lagrangian approach to parameterise this novel physics, as well as the parameter space that is compatible with the $S$, $T$, and $U$ parameters following the CDF-II data. We focused our attention on the chiral $B-L$ model. Following that, we reviewed the DM constraints derived from Planck's measurement of the relic density, as well as the constraints derived from the direct detection experiments. Finally, we demonstrated that, in the chiral $B-L$ model, the parameter space we found agrees with the current measurements of the $W$ boson mass, relic abundance, and direct detection experiments.    
\section{Kinetic-Mixing and $W$ mass}
\label{sec:KM}
In this section, we outline the theoretical framework for kinetic mixing in the presence of an additional Abelian gauge symmetry. We then demonstrate how a deviation in the $Z$-boson mass through kinetic mixing can be translated to the $W$-boson mass. We note that a kinetic mixing can occur provided there are two or more field strength tensors $B^{\mu\nu}$ and $X^{\mu\nu}$
which are neutral under some gauge symmetry.  Thus in our case with the gauge group $SU(3)_c \otimes SU(2)_L \otimes U (1)_Y \otimes U (1)_{B-L}$ , the kinetic terms can be expressed as follows
\begin{align}
\mathcal{L}_{\rm Kinetic} = -\frac{1}{4}B^{\mu\nu}B_{\mu\nu} - \frac{1}{4}X^{\mu\nu}X_{\mu\nu} - \frac{\kappa}{2} B^{\mu\nu}X_{\mu\nu},
\label{eq:kinetic-mixing}
\end{align}
where $B_{\mu\nu}$ and $X_{\mu\nu}$ are the filed strength tensors of the gauge groups $U(1)_Y$ and $U(1)_X$, respectively. The requirement of positive kinetic energy implies that the kinetic coefficient $|\kappa|<1$. One can diagonalize the kinetic mixing term as follows
\begin{align}
 \begin{pmatrix}
  \tilde{B} \\
  \tilde{X}  \\ 
 \end{pmatrix}
=
\begin{pmatrix}
 1 & \kappa \\
 0 & \sqrt{1-\kappa^2} \\
\end{pmatrix}
\begin{pmatrix}
 B \\
 X \\
\end{pmatrix}.
\end{align}
Let's first determine the gauge boson mass spectrum, setting the kinetic mixing $\kappa=0$ to fix our notation. With the kinetic mixing $\kappa=0$, the covariant derivative can be defined as
\begin{align}
D_{\mu}=\partial_{\mu}-ig_{s}T^{a}G^{a}_{\mu}-igT^{a}W^{a}_{\mu}-ig'Y B_{\mu}-ig_{x}Y_{x}X_{\mu},
\label{eq:covariant}
\end{align}
where gauge coupling $g_{x}$ is a free parameter. In addition to SM Higgs doublet $\Phi$, one adds a scalar, $\chi$, singlet of the SM but charged under $U(1)_{B-L}$, that spontaneously breaks
the $B-L$ symmetry. In the minimal case, the $U(1)_{B-L}$ charge of $\chi$ is $q_\chi=2$. To determine the gauge boson mass spectrum, we have to expand the following scalar kinetic terms 
\begin{align}
\mathcal{L}_{s}=(D^{\mu}\Phi)^{\dagger}(D_{\mu}\Phi)+(D^{\mu}\chi)^{\dagger}(D_{\mu}\chi),
\end{align}
and have to replace the fields $\Phi$ and $\chi$ by the following expressions such as
\begin{align}
\Phi=\frac{1}{\sqrt{2}}\begin{pmatrix}
0\\
v_\Phi+R_1\end{pmatrix},\,\, \chi=\frac{1}{\sqrt{2}} (v_\chi+R_2).
\end{align}
With this above replacement, we can expand the scalar kinetic terms $(D^\mu \Phi)^\dagger (D_\mu \Phi)$ and $(D^\mu\chi)^\dagger (D_\mu\chi)$ as follows
\begin{align}
& (D^{\mu}\Phi)^{\dagger}(D_{\mu}\Phi)\equiv\frac{1}{2}\partial^{\mu}R_1\partial_{\mu}R_1+\frac{1}{8}(R_1+v_\Phi)^{2}\Big(g^{2}|W_{1}^{\mu}-iW_{2}^{\mu}|^{2}+(gW_{3}^{\mu}-g'B^{\mu})^{2}\Big),\\
& (D^{\mu}\chi)^{\dagger}(D_{\mu}\chi)\equiv\frac{1}{2}\partial^{\mu}R_2\partial_{\mu}R_2+\frac{1}{2}(R_2+v_\chi)^{2}(g_{1}^{'} X^{\mu})^{2},
\end{align}
where we have defined $g_{1}^{'}=g_{x} q_\chi$.
With this, the mass matrix of the neutral gauge bosons is given by
\begin{align}
\mathcal{L}_M=\frac{1}{2} V_0^T M_G^2 V_0 ,
\end{align}
where
\begin{align}
V_0^T= \begin{pmatrix}B_\mu & W_{3\mu} & X_\mu \end{pmatrix} \text{   and    }
M_G^2=
\begin{pmatrix}
\frac{1}{4}g'^{2} v_\Phi^2 &  -\frac{1}{4} g g' v_\Phi^2   &   0 \\
-\frac{1}{4} g g' v_\Phi^2  &   \frac{1}{4} g^2 v_\Phi^2  &  0  \\
0  &  0  &   g_1^{'2} v_\chi^2  
\end{pmatrix} .
\end{align}
In the kinetic term diagonalized basis $\tilde{V}_0^T=(\tilde{B}_\mu\,\,W_{3\mu}\,\,\tilde{X}_\mu)$, the mass matrix of the neutral gauge boson can be written as
\begin{align}
\mathcal{L}_M=\frac{1}{2} \tilde{V}_0^T S^T M_G^2  S \tilde{V}_0 = \frac{1}{2} \tilde{V}_0^T  \tilde{M}_G^2  \tilde{V}_0 ,
\end{align}
where
\begin{align}
S=\begin{pmatrix} 
1  &  0  & -\frac{\kappa}{\sqrt{1-\kappa^2}}  \\
0  & 1   & 0 \\
0  &  0  & \frac{1}{\sqrt{1-\kappa^2}}
\end{pmatrix},
\tilde{M}_G^2=S^T  M_G^2  S = \begin{pmatrix}
\frac{1}{4}g'^{2} v_\Phi^2 &  -\frac{1}{4} g g' v_\Phi^2   &   \frac{1}{4} g' \tilde{g}_t v_\Phi^2 \\
-\frac{1}{4} g g' v_\Phi^2  &   \frac{1}{4} g^2 v_\Phi^2  &  -\frac{1}{4} g  \tilde{g}_t v_\Phi^2  \\
\frac{1}{4} g'  \tilde{g}_t v_\Phi^2  &  -\frac{1}{4} g  \tilde{g}_t v_\Phi^2  &  \frac{1}{4}  \tilde{g}_t^2 v_\Phi^2+ g_1^{''2} v_\chi^2  
\end{pmatrix},
\end{align}
with $\tilde{g}_t=-\frac{g'\kappa}{\sqrt{1-\kappa^2}}$. Following linear combination of $\tilde{B}^{\mu}$, $W_{3}^{\mu}$ and $\tilde{X}^{\mu}$ gives definite mass eigenstates $A^{\mu}$, $Z^{\mu}$ and $Z^{'\mu}$,
\begin{align}
 \begin{pmatrix}
  \tilde{B}^{\mu}\\
  W_{3}^{\mu}\\
  \tilde{X}^{\mu}
 \end{pmatrix}
= \begin{pmatrix}
 \cos\theta_{w}  & -\sin\theta_{w}~\cos\theta &  \sin\theta_{w}~\sin\theta \\
 \sin\theta_{w}  & \cos\theta_{w} \cos\theta  &  -\cos\theta_{w} ~\sin\theta \\
 0                      & \sin\theta                        &  \cos\theta \\
\end{pmatrix}
\begin{pmatrix}
 A^{\mu}\\
 Z^{\mu} \\
 Z^{'\mu} \\
\end{pmatrix},
\label{eq:final-gauge-tranformation}
\end{align}
where
\begin{align}
 \text{tan} 2\theta=\frac{2\tilde{g}_t\sqrt{g^{2}+g'^{2}}}{\tilde{g}_t^{2}+16\left(\frac{v_\chi}{2v_\Phi}\right)^{2}g_{1}^{''2}-g^{2}-g'^{2}} \text{  with  } \tilde{g}_t=-\frac{g'\kappa}{\sqrt{1-\kappa^2}} \text{ and } g_1^{''}=\frac{g_1'}{\sqrt{1-\kappa^2}}.
\end{align}
Masses of physical gauge bosons $A$, $Z$ and $Z^{'}$ are given by,
\begin{align}\label{KMZ}
 M_A=0,\,\,M_{Z,Z^{'}}^{2}=\frac{1}{8}\left(Cv_\Phi^{2}\mp\sqrt{-D+v_\Phi^{4}C^{2}}\right),
\end{align}
where,
\begin{align}
 C=g^{2}+g'^{2}+\tilde{g}_t^{2}+16\left(\frac{v_\chi}{2v_\Phi}\right)^{2}g_{1}^{''2},
 \hspace{0.5cm}
 D=16 v_\Phi^{2}v_\chi^{2}(g^{2}+g'^{2})g_{1}^{''2}.
\end{align}
The covariant derivative with the kinetic mixing  can be expressed in terms of the orthogonal fields $\tilde{B}$ and $\tilde{X}$ as 
\begin{align}
D_{\mu}=\partial_{\mu}-ig_{s}T^{a}G^{a}_{\mu}-igT^{a}W^{a}_{\mu}-ig'Y \tilde{B}_{\mu}-i\left(g_{x}Y_{x}\frac{1}{\sqrt{1-\kappa^2}}-g' Y \frac{\kappa}{\sqrt{1-\kappa^2}}\right)\tilde{X}_{\mu}.
\end{align}\\

\underline{\bf $W$ mass:} Now, let's try to see whether one can explain the CDF-II anomaly considering only kinetic mixing and ignoring any other loop corrections due to the new neutral gauge boson $Z'$. Specifically, we consider that the shift in $W$ boson mass measured by CDF-II also modifies the $Z$ boson mass at the tree level as the $\rho$ parameter should be equal to one at tree-level. Further, we investigate whether new physics contribution through kinetic mixing is sufficient to reduce this change in $Z$ mass to the experimental value of $M_Z=91.1876$ GeV~\cite{ParticleDataGroup:2020ssz}.
The tree-level formula for the $W$ and $Z$ mass is given as follows 
\begin{equation}
M_{W}^2 =\frac{v_{\Phi}^{2}g^{2}}{4},~~~~~M_{Z}^2|_{\kappa=0}=\frac{v_{\Phi}^{2}}{4}\left[g^2+g'^2\right] .
\label{eq:ZWmass}
\end{equation}
Taking the CDF II measured $W$ mass, $M_{W}=80.4335 \pm 0.0094$~GeV~\cite{CDF:2022hxs} and using the PDG values for other input parameters, $\sin^{2}\theta_{w} = 0.23121 \pm 0.00004,~ G_{f}=1.1663787(6)\times 10^{-5}$ (GeV)$^{-2}$~\cite{ParticleDataGroup:2020ssz}, we calculated weak couplings that is consistent with CDF-II measured $W$ mass and then calculated the $Z$ mass. With this, the central value of the theoretically computed $Z$ mass is given as 
\begin{equation}
M_{Z}|_{\kappa=0}= 91.7345~ \text{GeV},
\label{eq:Mztheoretical}
\end{equation}
which is of course larger than the experimental value of $M_Z=91.1876$ GeV~\cite{ParticleDataGroup:2020ssz}.
Note that in $B-L$ model, $Z$ mass is affected by the presence of the kinetic mixing parameter $\kappa$, see Eq.~\eqref{KMZ}:
\begin{equation}
M_{Z}=f\left(g_x,M_{Z'},\kappa \right).
\end{equation}
\begin{figure}[ht]
   \centering
   \captionsetup{justification=raggedright}
  \includegraphics[width=0.49\textwidth]{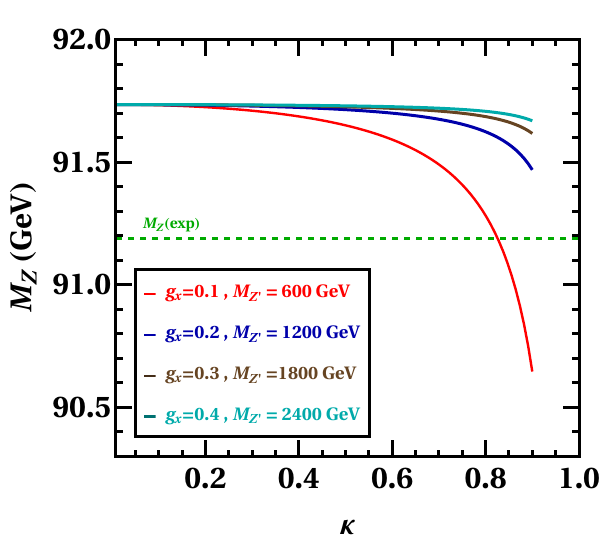}
  \includegraphics[width=0.49\textwidth]{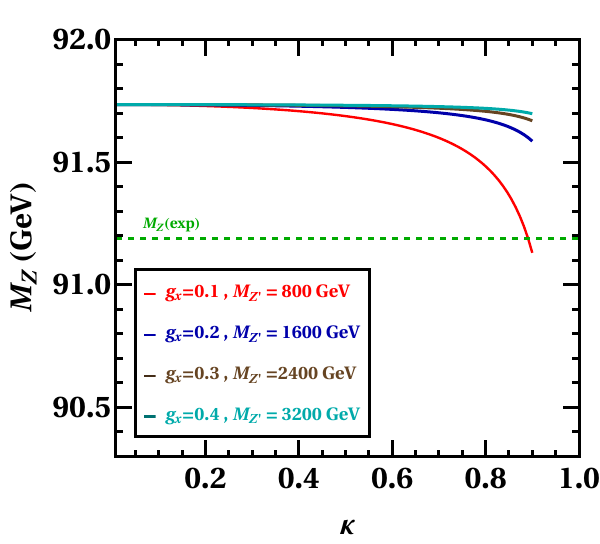}
  \includegraphics[width=0.49\textwidth]{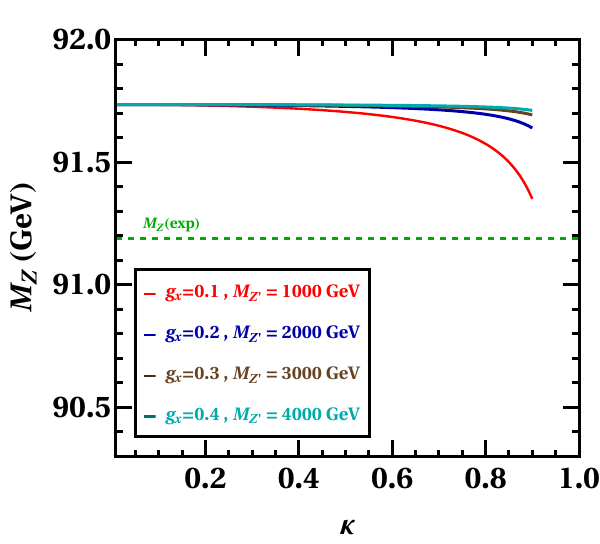}
  \caption{$Z$ mass versus kinetic mixing $\kappa$. The coloured lines in each panel correspond to different $g_{x}$ and $M_{Z'}$ values while keeping the ratio $\frac{M_{Z'}}{g_x}$ constant.}
  \label{fig:treelevel}
\end{figure}
As a result, the new physics contribution from kinetic mixing $\kappa$ can reduce the $Z$ mass to the experimental value. In Fig.~\ref{fig:treelevel}, we show how $Z$ mass depends on the kinetic mixing $\kappa$, $g_x$ and $M_{Z'}$. The various lines in each panel of Fig.~\ref{fig:treelevel} correspond to different values of $g_{x}$ and $M_{Z'}$ while the ratio $\frac{M_{Z'}}{g_x}$ remains constant. The ratio is kept at 6 TeV, 8 TeV and 10 TeV in the top left, top right and bottom panel, respectively. It is clear from Fig.~\ref{fig:treelevel} that when only the central values are considered, the change in $Z$ mass that touches the experimental value occurs only at low $Z'$ mass. Also, comparing the top left, top right and bottom panels of Fig.~\ref{fig:treelevel}, we see that at high $\frac{M_{Z'}}{g_{x}}$ ratio, the kinetic mixing is not sufficient to reduce $Z$ mass to the experimental value.\\
\begin{figure}[ht]
  \includegraphics[width=0.59\textwidth]{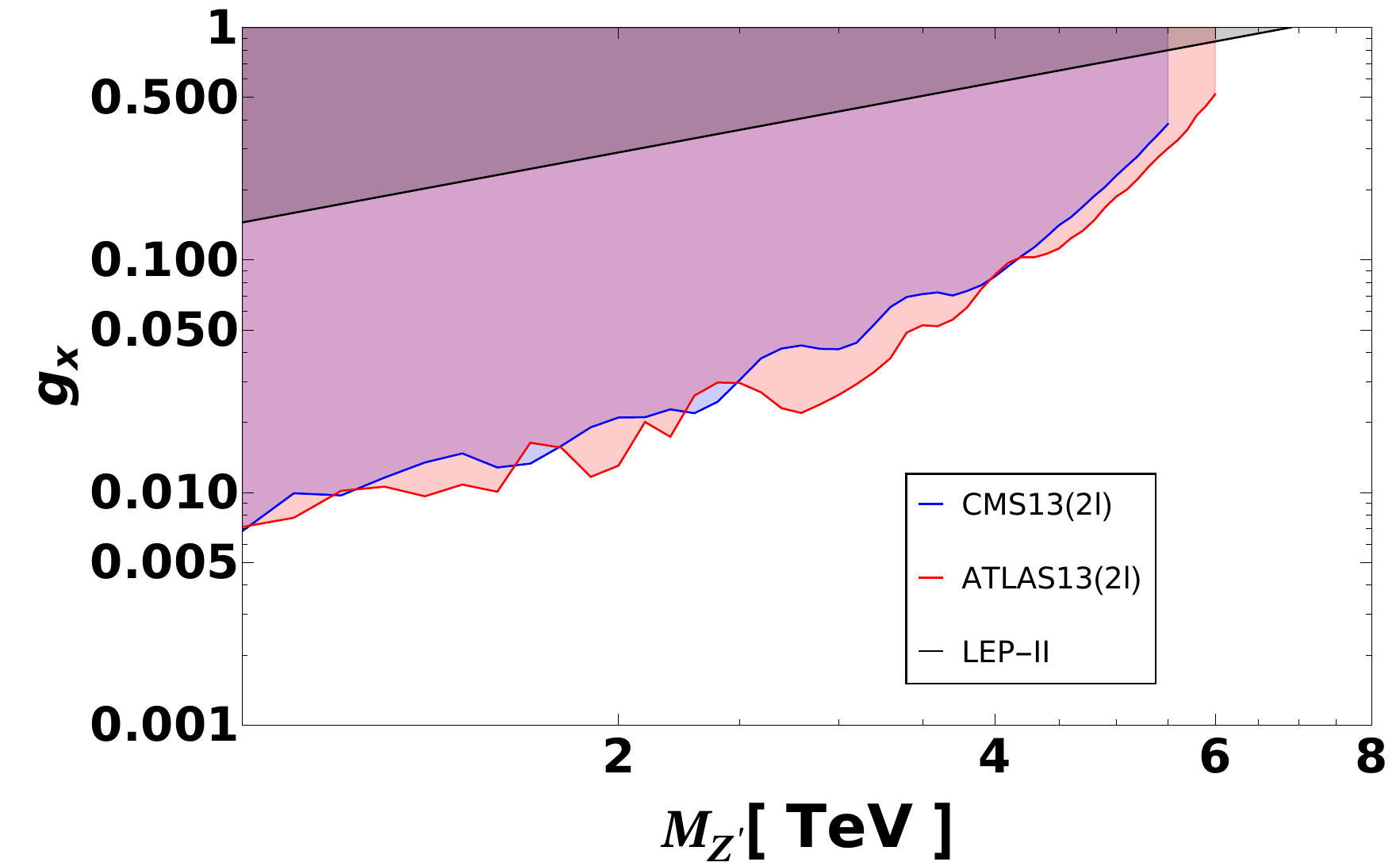}
  \caption{Constraint on $g_x$ as a function of $M_{Z'}$. The shaded regions are ruled out from LEP-II~\cite{Carena:2004xs,Cacciapaglia:2006pk}, ATLAS and CMS dilepton searches~\cite{ATLAS:2019erb,CMS:2021ctt}.}
  \label{fig:B-LConstraint}
\end{figure}
The mass of $Z'$ and the gauge coupling $g_x$ can be constrained with collider data. From LEP II data the bound
\begin{align}
\frac{M_{Z'}}{g_x}\gtrsim 6-7\,\text{TeV},
\end{align}
was derived in Refs.~\cite{Carena:2004xs,Cacciapaglia:2006pk,Das:2021esm}. Current ATLAS and CMS searches for dilepton resonances at the LHC can also be used to constrain $M_{Z'}$ via the Drell-Yan process, $pp\to Z'\to \ell\bar{\ell},\,\text{with}\,\ell=e,\mu$~\cite{ATLAS:2019erb,CMS:2021ctt,Das:2021esm}.  From Fig.~\ref{fig:B-LConstraint}, we see that the LHC dilepton constraints are the most stringent up to $M_{Z'} = 6$~TeV, beyond which the resonant $Z'$ production is kinematically limited at $\sqrt{s} = 13$~TeV. Hence, comparing Fig.~\ref{fig:treelevel} and Fig.~\ref{fig:B-LConstraint}, we can conclude that in view of current experimental constraints on $g_x-M_{Z'}$, it is not possible to explain CDF-II $W$ anomaly at tree-level with the help of kinetic mixing $\kappa$.
\section{Minimal and chiral $B-L$ model}
\label{sec:model}
We saw in the previous section that kinetic mixing alone is insufficient to explain the $W$ mass anomaly. From this point forward, we will ignore kinetic mixing and concentrate on the loop contribution from the $Z'$ gauge sector in order to explain the $W$ mass anomaly. In this section, we study two $U(1)_{B-L}$ gauged SM extensions: minimal and chiral extensions. Under $U(1)_{B-L}$, the SM quarks and leptons have charge $1/3$ and $-1$ respectively. As a result, $B-L$ is an anomalous symmetry that requires the inclusion of additional fermions to gauge it consistently. The gauge group $U(1)_{B-L}$ has the potential to cause the following triangle gauge anomalies: 
\begin{subequations}
\label{U1x anomaly cancellation}
\begin{align}
&[SU(3)_{c}]^2[U(1)_{B-L}]=  \sum\limits_{q} X_{q L} -  \sum\limits_{q} X_{q R} ,
\\& [SU(2)_{L}]^2[U(1)_{B-L}]=  \sum\limits_{l} X_{l L} + 3\sum\limits_{q} X_{q L} ,
\\& [U(1)_{Y}]^2 [U(1)_{B-L}] = \sum\limits_{l q} ( Y_{l L}^2   X_{l L} + 3 Y_{q L}^2   X_{q L}  )  - \sum\limits_{l q} ( Y_{l R}^2 X_{l R} + 3 Y_{q R}^2 X_{q R}  ),
\\& [U(1)_{Y}] [U(1)_{B-L}]^2 = \sum\limits_{l q} ( Y_{l L}   X_{l L}^2 + 3 Y_{q L}   X_{q L}^2  )  - \sum\limits_{l q} ( Y_{l R} X_{l R}^2 + 3 Y_{q R} X_{q R}^2  ) .
\end{align}
\end{subequations}  
In addition to this, we have two more equations
\begin{subequations}
\label{U1x anomaly cancellation1}
\begin{align}
&  [U(1)_{B-L}]^3= \sum\limits_{l q} ( X_{l L}^{3} + 3 X_{q L}^{3}  ) - \sum\limits_{l q} ( X_{l R}^{3} + 3 X_{q R}^{3}  ),
\\& [G]^2[U(1)_{B-L}]= \sum\limits_{l q} ( X_{l L} + 3 X_{q L}  ) - \sum\limits_{l q} ( X_{l R} + 3 X_{q R} ).
\end{align}
\end{subequations}
Where, $X$ is the $U(1)_{B-L}$ charge and $Y$ is the hyper charge. Anomalies from the first four equations of Eq.~\eqref{U1x anomaly cancellation} cancel within the SM particle content. To cancel anomalies arising from Eq.~\eqref{U1x anomaly cancellation1}, we add three generations of RHNs $(\nu_R^i, i = 1, 2, 3)$ with $U(1)_{B-L}$ charges $(x_{1},x_{2},x_{3})$. This gives us the following two conditions:
\begin{subequations}
\label{neutrino anomaly cancellation2}
\begin{align}
&  x_{1} + x_{2} + x_{3} = -3,\\
&  x_{1}^{3}+ x_{2}^{3} + x_{3}^{3} = -3 .
\end{align}
\end{subequations}
We will discuss two charge assignments for the $\nu_R^i$ that cancel anomalies. The first is the vector solution~(also sometimes called minimal $B-L$ extension), in which the RHNs have the same charge as the left-handed neutrino: $(-1,-1,-1)$. Second assignment makes neutrinos chiral under $U(1)_{B-L}$ with RHNs charges: $(5,-4,-4)$.
\begin{table}[ht]
\centering
\captionsetup{justification=raggedright}
 \begin{tabular}{|c|c|c|}
 \hline \centering
 ~~~Fields~~~ &   (  $SU(3)_{c} \otimes SU(2)_{L} \otimes U(1)_{Y} \otimes U(1)_{B-L}$  )    \\ 
 \hline 
 $L_{L}$  & ($1,2,-\frac{1}{2},-1$) \\ 
 \hline 
   $Q_{L}$  & ($3,2,\frac{1}{6},\frac{1}{3}$)  \\
   \hline 
 $e_{R}$  & ($1, 1, -1, -1$) \\ 
 \hline 
  \hline
 $\nu_{R}$  & ($1,1,0,-1$) \\ 
 \hline
 \hline 
 $u_{R}$ & ($3,1,\frac{2}{3},\frac{1}{3}$) \\ 
 \hline 
 $d_{R}$  & ($3,1,-\frac{1}{3},\frac{1}{3}$) \\ 
 \hline
 \hline 
 $\Phi$  & ($1,2,\frac{1}{2}, 0$) \\ 
 \hline 
 $\chi$  & ($1,1,0,2$) \\ 
 \hline 
 \end{tabular}  
\caption{Matter content and charge assignment of the vector $B-L$ model. For brevity, the generation index is suppressed.}
\label{tab:B-L}
\end{table}
\subsection{Vector $B-L$ Model}  
This model is a simple extension of the SM. The particle contents and their charges under the gauge group $SU(3)_{c} \otimes SU(2)_{L} \otimes U(1)_{Y} \otimes U(1)_{B-L}$ are given in Table~\ref{tab:B-L}. The new particles are three RHNs with $B-L$ charge $-1$ to cancel the gauge anomalies and a new scalar filed $\chi$, singlet of the SM but charged under $U(1)_{B-L}$, that spontaneously breaks the $B-L$ symmetry. We assign $B-L$ charge $+2$ for the scalar field $\chi$ so that $\nu_R^i$ gets Majorana mass after $B-L$ breaking, which further gives rise to light neutrino mass through the seesaw mechanism. We begin by writing down the Lagrangian of the scalar sector. The most general renormalizable and $SU(3)_{c} \otimes SU(2)_{L} \otimes U(1)_{Y} \otimes U(1)_{B-L}$ gauge-invariant scalar sector is given by  
\begin{align}
\mathcal{L}_s=(D^\mu\Phi)^{\dagger} (D_\mu\Phi) + (D^\mu\chi)^{\dagger} (D_\mu\chi) - \mathcal{V}(\Phi,\chi),
\end{align}
where the covariant derivative is defined in Eq.~\eqref{eq:covariant}. The scalar potential $\mathcal{V}(\Phi,\chi)$ is given by
\begin{align}
\mathcal{V}(\Phi,\chi) = m_{\chi}^2(\chi^{*}\chi) +\frac{1}{2}\lambda_{\chi}(\chi^{*}\chi)^2 + m_{\Phi}^{2}(\Phi^{\dagger}\Phi)  +\frac{1}{2}\lambda_{\Phi}(\Phi^{\dagger}\Phi)^2  + \lambda_{\Phi\chi}(\chi^{*}\chi)(\Phi^{\dagger}\Phi).
\label{eq:B-Lpotential}
\end{align}
The requirement to have a stable minimum for the potential at tree level implies the following conditions on quartic couplings
\begin{equation}
\lambda_{\Phi},\lambda_{\chi} > 0,~~~~~ \lambda_{\Phi \chi} > -\sqrt{\lambda_{\Phi}\lambda_{\chi}}\,.
\end{equation}
The breaking of the electroweak and the $U(1)_{B-L}$ gauge symmetries are driven by the vacuum expectation values (vev) of the scalar fields $\Phi$ and $\chi$. Denoting the vevs of field $\Phi$ and $\chi$ as $v_\Phi$ and $v_\chi$, the fields $\Phi$ and $\chi$ after symmetry breaking can be written in the following form:
\begin{equation}
\Phi=\frac{1}{\sqrt{2}} \begin{bmatrix}
\sqrt{2}G^{+} \\ v_{\Phi} + R_{1} + i I_{1}
\end{bmatrix}, \quad \quad \chi = \frac{1}{\sqrt{2}}(v_{\chi} + R_{2} +i I_{2}) .
\label{eq:vev-expansion}
\end{equation}
$G^{\pm}$ are the Goldstone boson corresponding to $W^{\pm}$. $I_1$ and $I_2$ will mix and give rise to the Goldstone bosons corresponding to the neutral gauge bosons $Z$ and $Z'$. The mass matrix of CP-even Higgs scalars in the basis $(R_1, R_2)$ reads as
\begin{align}
\mathcal{M}_R^2= \begin{pmatrix}
A  &  C \\
C  &   B
\end{pmatrix} =
\begin{pmatrix}
   v_{\Phi}^{2}\lambda_{\Phi} & v_{\Phi} v_{\chi} \lambda_{\Phi \chi}  \\
   v_{\Phi} v_{\chi} \lambda_{\Phi \chi} & v^{2}_{\chi} \lambda_{\chi}
\end{pmatrix}.
\label{eq:mass-matrix}
\end{align}
The mass eigenvalues of light and heavy mass eigenstates as
\begin{align}
m_h^2 & =\frac{1}{2}\left[A+B-\sqrt{(A-B)^2+4C^2}\right], \\
m_H^2 & =\frac{1}{2}\left[A+B+\sqrt{(A-B)^2+4C^2}\right].
\end{align}
We follow the convention $m_h^2 \leq m_H^2$ and have identified $h$ as the SM Higgs discovered at LHC, with mass $m_h=125$~GeV.
The two mass eigenstates $h, H$ are related with the $(R_1, R_2)$ fields through the following rotation matrix as
\begin{equation}
\begin{bmatrix}
h \\
H  \end{bmatrix} = U \begin{bmatrix}
R_1 \\
R_2 \end{bmatrix}
= \begin{bmatrix}
\cos\theta & -\sin\theta \\
\sin\theta & \cos\theta
\end{bmatrix} \begin{bmatrix}
R_1 \\
R_2
\end{bmatrix}, \,\, \text{with}\,\, \tan 2\theta=\frac{2C}{B-A}.
\end{equation} 
\par In the absence of kinetic mixing, neutral bosons cannot have mass mixing because the scalar doublet $\Phi$ does not carry any $B-L$ charge. The gauge boson masses are given as
\begin{equation}
M_{Z}^2=\frac{v_{\Phi}^{2}}{4}\left[g^2+g'^2\right],~~~~~M_{W}^2 =\frac{v_{\Phi}^{2}g^{2}}{4},~~~~~ M_{Z'} =2v_{\chi}g_{x},
\end{equation}
where $g$ and $g'$ are $SU(2)$ and hypercharge coupling respectively.\\
 \begin{figure}[ht]
  \includegraphics[width=0.5\textwidth]{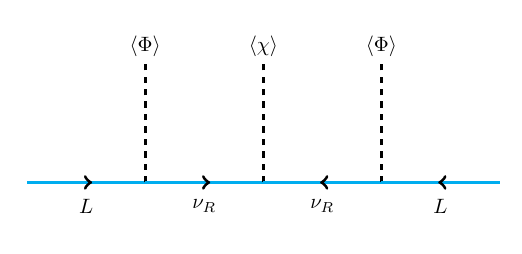}
  \caption{Neutrino mass generation in $B-L$ model through type-I seesaw mechanism.}
  \label{Feynman view}
\end{figure}
In the $B-L$ model, neutrino masses are generated by the seesaw mechanism as shown in Figure \ref{Feynman view}. Apart from the SM part, the Yukawa sector of the model can be written in a gauge-invariant way as
\begin{align}
-\mathscr{L}_{Y}& \supset Y_{\nu}^{ij} \overline{L}^{i}\tilde{\Phi} \nu^{j}_{R} + \frac{y_M^{ij}}{2}\overline{\nu^{c}_{R_{i}}}\nu_{R_{j}}\chi + \text{H.c.} ,
\label{eq:B-Lyukawa}
\end{align} 
The first and second terms will give the Dirac and Majorana contributions to the neutrino mass generation. $\chi$ gets vev and breaks the $B-L$ symmetry by two units; as a result, neutrinos gain majorana mass. We assume, without loss of any generality, a basis in which $y_M^{ij}$ is diagonal. After the breaking of electroweak and $U(1)_{B-L}$ symmetry, we can write the mass term as
\begin{align}
-\mathscr{L}_{M} \supset \overline{\nu_{L}}m_{D}\nu_{R} + \frac{1}{2}\overline{\nu^{c}_{R}}M_R\nu_{R} + \text{H.c.},
\end{align}
where $m_D=\frac{y_\nu v_\Phi}{\sqrt{2}}$ and $M_R=\frac{y_M v_\chi}{\sqrt{2}}$. Now using the fact that Majorana mass terms are symmetric and $\overline{\nu^{c}_{R}}m^{T}_{\nu}\nu^{c}_{L}=\overline{\nu_{L}}m_{\nu}\nu_{R}$, we can write the $\mathscr{L}_{M}$ in the following matrix form
\begin{align}
-\mathscr{L}_{M}\supset\quad \frac{1}{2}
\begin{pmatrix}
\overline{\nu_L}  &  \overline{(\nu_R)^c}
\end{pmatrix} \begin{pmatrix}
0 &  m_D \\
m_D^T & M_R
\end{pmatrix} \begin{pmatrix}
(\nu_L)^c \\
\nu_R
\end{pmatrix}.
\end{align}
From the above mass matrix, one can easily recover the seesaw formula for light Majorana neutrinos as, $\mathcal{M}_\nu\approx m_D M_R^{-1} m_D^T$ and the heavy neutrino mass as $M_N\approx M_R$ with the assumption $m_D\ll M_R$.
\begin{table}[ht]
 \begin{tabular}{|c|c|c|}
 \hline \centering
 ~~~Fields~~~ &  (  $SU(3)_{C} \otimes SU(2)_{L} \otimes U(1)_{Y} \otimes U(1)_{B-L}$  )   \\ 
 \hline
 $L_{L}$  & ($1,2,-1/2,-1$) \\ 
 \hline 
 $Q_{L}$  & ($3,2,1/6,1/3$) \\ 
 \hline 
 $e_{R}$  & ($1,1,-1,-1$) \\ 
 \hline 
 $u_{R}$  & ($3,1,2/3,1/3$) \\ 
 \hline 
 $d_{R}$  & ($3,1,-1/3,1/3$) \\ 
 \hline 
 \hline
 $\nu_{R}^1$  & ($1,1,0,5$) \\ 
 \hline 
 $\nu_{R}^{2,3}$  & ($1,1,0,-4$) \\ 
 \hline
  \hline 
 $\Phi$  & ($1,2,1/2,0$) \\ 
 \hline 
 $\varphi$  & ($1,2,1/2,-3$) \\  
  \hline 
 $\sigma$  & ($1,1,0,3$) \\  
 \hline 
 $\chi_{d}$  & ($1,1,0,1/2$) \\  
 \hline 
 \end{tabular}  
\caption{Matter content and charge assignment of the chiral $B-L$ model. For brevity, the generation index is suppressed.}
\label{tab:B-Lchiral}
\end{table}
\subsection{Chiral $B-L$ Model}
Another $U(1)_{B-L}$ gauged model will be discussed in this section. The particle contents and their charges under the gauge group $SU(3)_{C} \otimes SU(2)_L \otimes U(1)_Y \otimes U(1)_{B-L}$ are given in Table. \ref{tab:B-Lchiral}. For RHNs, we use a chiral anomaly cancellation solution $(5,-4,-4)$. Apart from the SM particle content and RHNs, in the scalar sector, we add one more $SU(2)_L$ doublet $\varphi$ and a scalar singlet $\sigma$. $\varphi$ is with hypercharge  $+1/2$ and $U(1)_{B-L}$ charge $-3$, whereas scalar $\sigma$ has $U(1)_{B-L}$ charge $+3$. We also include a scalar Dark Matter (DM) particle $\chi_d$ with a charge of $U(1)_{B-L}$ of $+1/2$. The advantage is that one does not need to impose any ad hoc $\mathbb{Z}_2$ symmetry to stabilise the DM. Instead, the stability of $\chi_d$ can be guaranteed by this nontrivial $B-L$ charge. The most general renormalizable and $SU(3)_c\otimes SU(2)_L\otimes U(1)_Y\otimes U(1)_{B-L}$ gauge-invariant scalar sector is given by
\begin{align}
\mathcal{L}_s=(D_\mu\Phi)^{\dagger} D_\mu\Phi + (D_\mu\varphi)^{\dagger} D_\mu\varphi + (D_\mu\sigma)^{\dagger} D_\mu\sigma + (D_\mu\chi_d)^{\dagger} D_\mu\chi_d - \mathcal{V}(\Phi,\varphi,\sigma,\chi_d), 
\end{align}
where again the covariant derivative is defined in Eq.~\eqref{eq:covariant}. The scalar potential $\mathcal{V}(\Phi,\varphi,\sigma,\chi_d)$ is given by,
\begin{align}
\label{eq:chiral-potential}
\begin{split}
\mathcal{V}(\Phi,\varphi,\sigma,\chi_d)& = m_{\sigma}^2(\sigma^{*}\sigma) +\frac{1}{2}\lambda_{\sigma}(\sigma^{*}\sigma)^2 + m_{\Phi}^{2}(\Phi^{\dagger}\Phi)  +\frac{1}{2}\lambda_{\Phi}(\Phi^{\dagger}\Phi)^2 
 +  m_{\varphi}^{2}(\varphi^{\dagger}\varphi)  +\frac{1}{2}\lambda_{\varphi}(\varphi^{\dagger}\varphi)^2 \\& 
~~ + m_{\chi_{d}}^2(\chi_{d}^{*}\chi_{d}) +\frac{1}{2}\lambda_{\chi_{d}}(\chi_{d}^{*}\chi_{d})^2 - \mu (\Phi^{\dagger} \varphi) \sigma - \mu (\varphi^{\dagger}\Phi)\sigma^{*} + \lambda_{\Phi\sigma} (\Phi^{\dagger}\Phi)(\sigma \sigma^{*}) \\&~~+ \lambda_{\varphi\sigma} (\varphi^{\dagger}\varphi)(\sigma \sigma^{*})+\lambda_{\Phi \varphi_{1}} (\Phi^{\dagger}\Phi)(\varphi^{\dagger}\varphi) + \lambda_{\Phi \varphi_{2}} (\Phi^{\dagger}\varphi)(\varphi^{\dagger}\Phi) + \lambda_{\Phi \chi_{d}} (\Phi^{\dagger}\Phi)(\chi_{d}^{*}\chi_{d}) \\&~~+ \lambda_{\varphi \chi_{d}}(\varphi^{\dagger}\varphi)(\chi_{d}^{*}\chi_{d}) + \lambda_{\sigma \chi_{d}} (\sigma \sigma^{*})(\chi_{d}^{*} \chi_{d})    . 
\end{split}
\end{align}

The necessary and sufficient conditions for tree level vacuum stability can be obtained by examining the copositivity of the matrix of quartic couplings, $\Lambda$ \cite{Kannike:2012pe}. To construct this matrix, we have parameterized the field bilinears as,
\begin{equation}
\Phi^{\dagger}\Phi = h_{\Phi}^{2},\,\varphi^{\dagger}\varphi = h_{\varphi}^{2},\, \chi_{d}^{\dagger}\chi_{d} = h_{\chi_{d}}^{2},\, \sigma^{\dagger}\sigma = h_{\sigma}^{2},\, \Phi^{\dagger}\varphi = h_{\Phi}h_{\varphi}\rho e^{i \theta} .
\end{equation} 
The parameter $\rho$ takes values in the interval $|\rho| \in [0,1]$. After minimizing the potential with respect to $\rho$ and $\theta$, the matrix $\Lambda$, in the basis $(h_{\Phi}^{2}, h_{\varphi}^{2}, h_{\sigma}^{2}, h_{\chi_{d}}^{2})$ is given by \cite{Kannike:2012pe},
\begin{equation}
2 \Lambda =
\begin{bmatrix}
 \lambda_{\Phi} &  \lambda_{\Phi \varphi_{1}} + \text{min}(0,\lambda_{\Phi \varphi_{2}}) & ~~~  \lambda_{\Phi \sigma} & ~~~ \lambda_{\Phi \chi_{d}} ~~~ \\
 \lambda_{\Phi \varphi_{1}} +\text{min}(0,\lambda_{\Phi \varphi_{2}}) & \lambda_{\varphi} & ~~~ \lambda_{\varphi \sigma} & \lambda_{\varphi \chi_{d}} \\
 \lambda_{\Phi \sigma} & \lambda_{\varphi \sigma} & ~~\lambda_{\sigma} & \lambda_{\sigma \chi_{d}} \\
 \lambda_{\Phi \chi_{d}} & \lambda_{\varphi \chi_{d}} & ~~~\lambda_{\sigma \chi_{d}} & \lambda_{\chi_{d}}
\end{bmatrix}
\end{equation}
Depending on the sign distribution of the off-diagonal elements, the conditions for the copositivity of this fourth-order matrix can be classified into eight separate cases \cite{Chakrabortty:2013mha}. Here in this work, for copositivity, we consider all quartic couplings to be positive \cite{Chakrabortty:2013mha}.

Neutral components of $\Phi$ and $\varphi$ spontaneously break electroweak symmetry. A singlet scalar $\sigma$, along with $\varphi$, breaks the $U(1)_{B-L}$ spontaneously.
First we solve the minimization equations for the mass parameters $m_{\Phi}, m_{\varphi}, m_\sigma$ in the potential. We get
\begin{subequations}
\label{tedpole 2}
\begin{align}
&  2m_{\Phi}^{2}+ v_{\Phi}^{2}\lambda_{\Phi} - \frac{\sqrt{2}\mu}{v_{\Phi}}v_{\varphi} v_{\sigma} + v_{\chi}^{2}\lambda_{\Phi\sigma} + v_{\varphi}^{2}(\lambda_{\Phi \varphi_{1}} + \lambda_{\Phi \varphi_{2}} ) = 0,\\&
2m_{\sigma}^{2} - \frac{\sqrt{2}\mu}{v_{\sigma}}v_{\Phi}v_{\varphi} + v_{\sigma}^{2}\lambda_{\sigma} + v_{\Phi}^{2}\lambda_{\Phi \sigma}+v_{\varphi}^{2}\lambda_{\varphi \sigma} =0 ,\\&
2m_{\varphi}^{2} -\frac{\sqrt{2} \mu}{v_{\varphi}}v_{\Phi}v_{\sigma} + v_{\sigma}^{2}\lambda_{\varphi \sigma} + v_{\varphi}^{2}\lambda_{\varphi}  + v_{\Phi}^{2}(\lambda_{\Phi \varphi_{1}} + \lambda_{\Phi \varphi_{2}})=0.
\end{align}
\end{subequations}
The fields $\Phi$, $\varphi$ and $\sigma$ can be written in the unitary gauge after symmetry breaking in the following form: 
\begin{equation}
\label{kin field expansion around minima2}
\Phi=\frac{1}{\sqrt{2}} \begin{bmatrix}
\sqrt{2}G^{+}_{1} \\ v_{\Phi} + R_{1} + i I_{1}
\end{bmatrix} ,~~\varphi=\frac{1}{\sqrt{2}} \begin{bmatrix}
\sqrt{2}G^{+}_{2} \\ v_{\varphi} + R_{2} + i I_{2}
\end{bmatrix},~ \sigma = \frac{1}{\sqrt{2}}(v_{\sigma} + R_{3} +i I_{3}) .
\end{equation}
$G_{1}^{\pm}$ and $G_{2}^{\pm}$  will mix and give rise to the Goldstone boson $G^{\pm}$ corresponding to the $W^{\pm}$ boson. One electrically charged field remains as the physical field. The mass matrix of these electrically charged fields in the basis $(G_{1}^{+}, G_{2}^{+})$ reads as
\begin{equation}
\mathcal{M}_{\pm}^2 = \frac{1}{2}
\begin{pmatrix}
\frac{\sqrt{2}\mu v_{\sigma}v_{\varphi}}{v_{\Phi}}-v_{\varphi}^{2}\lambda_{\Phi \varphi_{2}} ~&~ v_{\Phi}v_{\varphi}\lambda_{\Phi \varphi_{2}} - \sqrt{2}\mu v_{\sigma}\\\\
 v_{\Phi}v_{\varphi}\lambda_{\Phi \varphi_{2}} - \sqrt{2}\mu v_{\sigma} ~&~ \frac{\sqrt{2}\mu v_{\sigma}v_{\Phi}}{v_{\varphi}}-v_{\Phi}^{2}\lambda_{\Phi \varphi_{2}} 
\end{pmatrix}.
\end{equation}
Mass eigen states are given as
\begin{equation}
M_{H^{\pm}}^{2} = \frac{v^{2}}{2v_{\Phi}v_{\varphi}} \left( \sqrt{2}\mu v_{\sigma} -v_{\Phi}v_{\varphi}\lambda_{\Phi\varphi_{2}}        \right),
\end{equation}
where, $v = \sqrt{v_{\Phi}^{2} + v_{\varphi}^{2}} $.
\par The two mass eigenstates $G^{\pm}, H^{\pm}$ are related with the $(G_{1}^{\pm}, G_{2}^{\pm})$ fields through the following rotation matrix as
\begin{equation}
\begin{bmatrix}
G^{\pm} \\
H^{\pm}  \end{bmatrix} = U \begin{bmatrix}
G_{1}^{\pm} \\
G_{2}^{\pm} \end{bmatrix}
= \begin{bmatrix}
\cos\alpha & \sin\alpha \\
-\sin\alpha & \cos\alpha
\end{bmatrix} \begin{bmatrix}
G_{1}^{\pm} \\
G_{2}^{\pm}
\end{bmatrix}, \,\, \text{with}\,\, \tan \alpha=\frac{v_{\varphi}}{v_{\Phi}}.
\end{equation}
In pseudo-scalar sector $I_{1}$, $I_{2}$ and $I_{3}$ mix together and gives two Goldstone boson $G_{1}^{0}$, $G_{2}^{0}$ corresponding to the neutral gauge bosons $Z$ and $Z'$, and one pseudo scalar field remains as a physical massive field $H^{0}$. The mass matrix in the basis $(I_{1},I_{2},I_{3})$ can be written as
\begin{equation}
\mathcal{M}_{I}^2=\frac{1}{\sqrt{2}}
\begin{pmatrix}
\frac{\mu v_{\varphi} v_{\sigma}}{v_{\Phi}}~&~ -\mu v_{\sigma} ~&~ -\mu v_{\varphi}\\
-\mu v_{\sigma} ~&~ \frac{\mu v_{\Phi} v_{\sigma}}{v_{\varphi}} ~&~ \mu v_{\Phi}\\
-\mu v_{\varphi}  ~&~ \mu v_{\Phi} ~&~ \frac{\mu v_{\Phi} v_{\varphi}}{v_{\sigma}} 
\end{pmatrix}.
\end{equation}
Mass of the physical eigenstate is given as
\begin{equation}
M_{H^{0}}^{2} = \frac{\mu}{\sqrt{2}v_{\Phi}v_{\varphi}v_{\sigma}}\left( v_{\Phi}^{2}v_{\varphi}^{2} + v_{\sigma}^{2}v^{2} \right).
\label{eq:mh0}
\end{equation}
\par Mass eigenstates $G_{1}^{0},G_{2}^{0},H^{0}$ are related with the $(I_{1},I_{2},I_{3})$ fields through the following rotation matrix as
\begin{equation}
\begin{bmatrix}
G_{1}^{0} \\
G_{2}^{0}\\
H^{0}  \end{bmatrix} = U \begin{bmatrix}
I_{1} \\
I_{2} \\
I_{3} \end{bmatrix}
= \begin{bmatrix}
\cos\alpha & \sin\alpha & 0 \\
-\sin\alpha \cos\beta & \cos\alpha \cos\beta & -\sin\alpha \\
-\sin\alpha \sin\beta & \cos\alpha \sin\beta & \cos\beta
\end{bmatrix} \begin{bmatrix}
I_{1} \\
I_{2} \\
I_{3}
\end{bmatrix},
\end{equation}
Where,
\begin{equation}
\tan\alpha = \frac{v_{\varphi}}{v_{\Phi}},~ \tan\beta = \frac{v_{\sigma}v}{v_{\Phi}v_{\varphi}}.
\end{equation}
Three CP-even neutral scalars are mixed together. The mass matrix in the basis $(R_{1},R_{2},R_{3})$ can be expressed as
\begin{equation}
\mathcal{M}_{S}^2=
\frac{1}{2}
\begin{bmatrix}
 2v_{\Phi}^{2}\lambda_{\Phi}+\frac{\sqrt{2}\mu}{v_{\Phi}}v_{\varphi}v_{\sigma}  &~~  2v_{\Phi}v_{\varphi}\lambda_{12} - \sqrt{2}\mu v_{\sigma}  &~~   2v_{\Phi}v_{\sigma}\lambda_{\Phi \sigma} -\sqrt{2}v_{\varphi}\mu \\\\
  2v_{\Phi}v_{\varphi}\lambda_{12} -\sqrt{2}\mu v_{\sigma}  &~~  2v_{\varphi}^{2}\lambda_{\varphi}+ \frac{\sqrt{2} \mu}{v_{\varphi}}v_{\Phi}v_{\sigma}    &~~  2 v_{\varphi}v_{\sigma}\lambda_{\varphi \sigma}- \sqrt{2}v_{\Phi}\mu\\\\
 2 v_{\Phi}v_{\sigma}\lambda_{\Phi\sigma}-\sqrt{2}v_{\varphi}\mu  &~~   2v_{\varphi}v_{\sigma}\lambda_{\varphi \sigma}  - \sqrt{2}v_{\Phi}\mu &~~  2v_{\sigma}^{2}\lambda_{\sigma} +\frac{\sqrt{2}\mu}{v_{\sigma}}v_{\Phi}v_{\varphi}, \\
\end{bmatrix},
\end{equation}
where, $\lambda_{12}=\lambda_{\Phi \varphi_{1}} + \lambda_{\Phi \varphi_{2}}~.$
The matrix $\mathcal{M}^2_{S}$ can be diagonalized by an orthogonal matrix : $\mathcal{O}_{R}^T M_{R}^2 \mathcal{O}_R = \text{diag}(m_{H_1}^2,m_{H_2}^2,m_{H_3}^2)$ with
\begin{eqnarray}
 \label{rot1}
  \left( \begin{array}{c} H_1\\ H_2\\ H_3\\ \end{array} \right) = 
 \mathcal{O}_{R} \left( \begin{array}{c} R_1\\ R_2\\ R_3\\ 
 \end{array} \right). 
 \end{eqnarray}
We assume the mass eigenstates to be ordered by their masses $m_{H_1}\leq m_{H_2}\leq m_{H_3}$. $H_{1}=h$ is identified with the SM Higgs of $125$ GeV. We will use the standard parameterization $\mathcal{O}_{R} = R_{23} R_{13} R_{12}$ where
\begin{equation}
R_{12} = \left(
\begin{array}{ccc}
c_{12} & -s_{12} & 0\\
s_{12} & c_{12} & 0\\
0 & 0 & 1
\end{array} \right), 
\quad R_{13} = \left(
\begin{array}{ccc}
c_{13} & 0 & -s_{13}\\
0 & 1 & 0\\
s_{13} & 0 & c_{13}
\end{array} \right), 
\quad R_{23} = \left(
\begin{array}{ccc}
1 & 0 & 0\\
0 & c_{23} &  -s_{23}\\
0 & s_{23} & c_{23}
\end{array} \right)
\end{equation}
$c_{ij} = \cos \theta_{ij}, s_{ij} = \sin \theta_{ij}$, where the angles $\theta_{ij}$ can be chosen to lie in the range $-\frac{\pi}{2}\leq\theta_{ij}\leq \frac{\pi}{2}$. Finally, the mass of the DM $\chi_d$  will be given as
\begin{equation}\label{DMmass}
M_{\rm DM}^{2} = \frac{2m_{\chi_{d}}^{2}+v_{\Phi}^{2} \lambda_{\Phi \chi_{d}} + v_{\sigma}^{2} \lambda_{\sigma \chi_{d}} + v_{\varphi}^{2} \lambda_{\varphi \chi_{d}}}{2}.
\end{equation}
\underline{\bf Neutrino mass:} The Yukawa sector of the model can be written in a gauge-invariant way as
\begin{equation}
\begin{split}
-\mathscr{L}_{Y}& = Y_{e}^{ij}\overline{L}^{i} \Phi e_{R}^{j} +Y_{u}^{ij} \overline{Q}^{i} \tilde{\Phi} u_{R}^{j}  + Y_{d}^{ij}\overline{Q}^{i} \Phi d_{R}^{j} + Y_{\nu}^{ij}\overline{L}^{i}\tilde{\varphi}\nu_{R}^{j}  +  \text{H.c.}
\end{split} 
\label{eq:Yukawa}
\end{equation}
 \begin{figure}[ht]
   \centering
   \captionsetup{justification=centering}
  \includegraphics[width=0.35\textwidth]{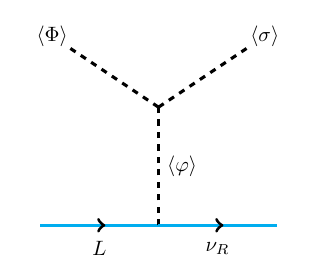}
  \caption{Neutrino mass generation in Chiral $B-L$ model through Dirac type II seesaw}\label{Feynman view 2}
\end{figure}
We see from the last term of Eq.~\eqref{eq:Yukawa} that the two RHNs pair up with the two left-handed neutrinos of the SM to form Dirac particles. Note the importance of unconventional $B-L$ charges of $\varphi$ and $\nu_R^i$ to generate the Dirac neutrino mass. This $B-L$ charge assignment is needed to forbid Majorana mass terms for the $\nu_R^i$ while simultaneously enforcing a Yukawa coupling structure in which only $\varphi$ couples to RHNs as shown in the Figure \ref{Feynman view 2}. 
As $\varphi$ gets vev, it breaks the lepton number by 3 units and ensures the Dirac nature of the neutrinos \cite{Hirsch:2017col,Ma:2014qra}. After the spontaneous breaking of electroweak and $U(1)_{B-L}$ symmetry, we can write the neutrino mass term as
\begin{equation}
-\mathscr{L}_{M} =  \overline{\nu_{L}} m_{\nu} \nu_{R}+\text{H.c.}
\end{equation}
where $m_{\nu}=\frac{Y_{\nu}^{ij} v_\varphi}{\sqrt{2}}$. The smallness of the neutrino masses relative to those of the quarks and charged leptons is explained by the smallness of the second Higgs doublet vev $v_{\varphi}\sim \text{eV}$ for large Yukawa coupling $Y_\nu\sim\mathcal{O}(1)$. In this model, smallness of $v_{\varphi}$ arises very naturally and this can be understood from Eq.~\eqref{eq:mh0} with the approximation $v_\sigma\gg v_{\Phi,\varphi}$:
\begin{align}
v_{\varphi}\approx \frac{\mu v_\sigma v^2}{\sqrt{2} M_{H^0}^2 v_\Phi}.
\end{align}
Hence, the vev of the neutral component of the field $\varphi$ is inversely proportional to the mass of the heavy scalar. This provided a natural explanation for the low vev and, thus, low neutrino masses. Note that, due to the chiral anomaly cancellation condition for RHNs, the Yukawa term can only give masses to two of the neutrinos. One can add another doublet to give mass to the third neutrino \cite{Ma:2015raa}. For simplicity, we assumed one neutrino to be massless.\\\\
\underline{\bf Gauge sector:} As field $\varphi$ is charged under both SM and $U(1)_{B-L}$ gauge group, even in the absence of kinetic mixing, this will introduce mixing between SM neutral boson $Z$ with the new neutral gauge boson $Z'$ corresponding to $U(1)_{B-L}$, at tree level. Again, to determine the gauge boson mass spectrum, we have to expand the following kinetic terms:
\begin{align}
(D_\mu\Phi)^{\dagger} D_\mu\Phi + (D_\mu\varphi)^{\dagger} D_\mu\varphi + (D_\mu\sigma)^{\dagger} D_\mu\sigma ,
\end{align}
and have to replace the fields $\Phi$, $\varphi$ and $\sigma$ by the expressions given in Eq.~\eqref{kin field expansion around minima2}. The gauge bosons mass matrix in the basis $(B^{\mu},W_{3}^{\mu},X^{\mu})$ can be written as
\begin{equation}
\label{eq:gauge-boson-mass} \mathcal{M}_V^2 =
\frac{v^{2}}{4} \begin{bmatrix}
g'^{2} &~~ -gg'&~~ -6u^{2}g'g_{x} \\
-gg'&~~  g^{2} &~~ 6u^{2}gg_{x}\\
-6u^{2}g'g_{x} &~~ 6u^{2}gg_{x} &~~ 36b^{2}g_{x}^{2}
\end{bmatrix},\,\, \text{where}\,\, u=\frac{v_{\varphi}}{v},\text{ and } b^{2}= u^{2} + \frac{v_{\sigma}^{2}}{v^{2}}.
\end{equation}
Mass matrix in Eq.~\eqref{eq:gauge-boson-mass} can be diagonalized by the following unitary matrix
\begin{equation}
\label{unitary matrix}
\begin{bmatrix}
A^{\mu} \\
Z^{\mu} \\
Z^{\prime\mu}
\end{bmatrix} =
\begin{bmatrix}
\cos\theta_{w} &~ \sin\theta_{w} &~0\\
-\cos\alpha' \sin\theta_{w} & \cos\alpha' \cos\theta_{w}
&~ -\sin \alpha'\\
-\sin\alpha' \sin\theta_{w} &~  \sin\alpha'\cos\theta_{w} &~ \cos\alpha' 
\end{bmatrix}  \begin{bmatrix}
B^{\mu} \\
W_{3}^{\mu}\\
X^{\mu}
\end{bmatrix},
\end{equation}
where, $\tan\theta_{w} = \frac{g'}{g}$, and $\tan2\alpha' = \frac{C'}{B'}$.
After rotation, we get a massless photon and two heavy bosons:
\begin{equation}
M_{A}=0,\,\, M_{Z}^{2} = \frac{v^{2}}{8}\left( A' -\sqrt{B'^{2} + C'^{2}}\right) \,\,\text{and}\,\, M_{Z'}^{2} = \frac{v^{2}}{8}\left( A' +\sqrt{B'^{2} + C'^{2}}\right),
\end{equation}
where, $A'=36b^{2}g_{x}^{2} + (g^{2} + g'^{2}),~~B'=36b^{2}g_{x}^{2} -(g^{2} + g'^{2})$ and $C'=12g_{x}u^{2}\sqrt{g^{2} + g'^{2}}$.

In models with an extended Higgs sector, there is an alignment limit where the lightest CP-even Higgs boson exhibits interactions identical to those of the SM Higgs boson \cite{Branco:2011iw}. In chiral $B-L$ model the coupling of lightest CP even scalar $h$ with SM gauge bosons and charged fermions are given as
\begin{equation}
\label{Hcoupl}
\begin{aligned}
& \frac{g^{2}}{2} (v_{\Phi} c_{13}c_{12}-v_{\varphi}c_{13}s_{12})hW^{\pm , \mu}W_{\mu}^{\pm} + \frac{g^2 + g'^{2}}{4} (v_{\Phi} c_{13}c_{12}-v_{\varphi}c_{13}s_{12})(\cos^{2}\alpha') h Z^{\mu}Z_{\mu}  \\ \nonumber & \hspace{9cm}+ \left(\frac{m_{ij}}{v_{\Phi}}c_{13}c_{12} \right)h\bar{f_{i}}f_{j} .
\end{aligned}
\end{equation}
%
As vev of the second Higgs doublet is small, $ v_{\varphi} \sim \text{eV} $. Consequently, $ v = \sqrt{v_{\Phi}^{2} + v_{\varphi}^{2}} \approx v_{\Phi} = 246.22 \, \text{GeV} $. This implies that, for the alignment limit, a sufficient condition is that the gauge eigenstate $ R_{1} $ remains adequately decoupled from $ R_{2} $ and $ R_{3} $ ($ c_{12} \approx c_{13} \approx 1 $), and the neutral gauge boson mixing angle $ \alpha' $ remains sufficiently small. As we choose the lightest CP-even scalar, $h$, to represent the SM Higgs, we will work within this alignment limit. 
\section{$W$ mass and the $S$, $T$, $U$ Parameters}
\label{sec:wstu}
In the SM, the $W$ boson mass can be calculated very precisely in terms of the precisely measured input parameters $\{G_F, \alpha_{\text{em}}, M_Z\}$. The $W$ boson mass is related with these parameters in the following way~\cite{Lopez-Val:2014jva,Hollik:2006hd,Amrith:2018yfb}:
\begin{equation}
M_{W}^{2} =\frac{M_{Z}^{2}}{2} \left[1+ \sqrt{1-\frac{4\pi \alpha_{\rm em}}{\sqrt{2} G_{F} M_{Z}^{2}\left( 1+ \Delta_{r}   \right) }}  ~ \right],
\label{eq:MW}
\end{equation}
where $\Delta_r$ represents the quantum corrections. Taking the central values of the input parameters, $M_Z = 91.1876$ GeV, $\alpha_{\rm em}^{-1} = 137.036$, $G_F = 1.1663787 \times 10^{-5}$ GeV$^{-2}$ and considering the SM value of $\Delta_{r} \approx 0.038$, Eq. (\ref{eq:MW}) gives us the theoretical prediction of $W$ boson mass $80.360$ GeV, with a theoretical uncertainty of near about $4$ MeV. This theoretical prediction is $7 \sigma$ away from the recently announced CDF-II results. Note that the new physics contribution to the parameter $\Delta_{r}$ can be reparametrised in terms of the self-energy corrections to the gauge bosons. Specifically, dominant BSM effects can be written in terms of the three gauge boson self-energy parameters known as the oblique parameters $S$, $T$, and $U$ provided that the new physics mass scale is greater than the electroweak scale and that it contributes only through virtual loops to the electroweak precision observables. The $W$ boson mass in terms of these parameters can be written as \cite{Cai:2016sjz}:
\begin{equation}
M_{W}=M_{W}^{\rm SM} \left[  1- \frac{\alpha_{em}}{4(\cos^{2}\theta_{w} - \sin^{2}\theta_{w})}\left(  S-1.55T-1.24U   \right)      \right].
\end{equation} 
Recently, Ref.~\cite{Lu:2022bgw} gave the values of these parameters from an analysis of precision electroweak data including the CDF-II new result of the $W$-mass at $1\sigma$:
\begin{equation}
\label{STUdata}
S=0.06 \pm 0.10,~ T=0.11 \pm 0.12,~ U= 0.14 \pm 0.09.
\end{equation}
with the correlation
\begin{equation}
\rho_{ST} = 0.90,~  \rho_{SU}= -0.59~ \text{and} ~\rho_{TU} = -0.85.
\end{equation}
\section{New Physics Contribution To ~$S$, $T$, $U$}
\label{sec:STUspace}
Using six dimensional $SU(2)_{L}$ invariant effective operator, we can parametrise new physics that only couples to SM vector bosons and Higgs. Effects related with dimension 6 operators can be expressed in the following way~\cite{Barbieri:2004qk,Cacciapaglia:2006pk}
\begin{equation}
\mathscr{L} = \mathscr{L}_{SM} +\frac{2}{v^{2}} \left( c_{WB}O_{WB} + c_{H}O_{H} + c_{WW}O_{WW} + c_{BB}O_{BB} \right).
\end{equation}
 \begin{figure}[ht]
   \centering
   \captionsetup{justification=raggedright}
  \includegraphics[width=0.49\textwidth]{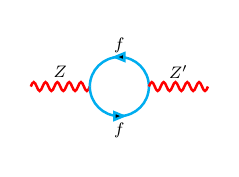}
  \caption{$Z$ and $Z'$ boson mixing at loop level in the vector B-L model.} \label{ZZP}
\end{figure}

We presented two generic models with heavy new neutral vector bosons in section~\eqref{sec:model}. In the first model, the SM doublet scalar is not charged under the new symmetry, and hence, there is no mass mixing between SM neutral bosons and the new $Z'$ boson at tree level. Since both leptons and quarks are  charged under the new
symmetry, loop effects can mix $Z$ and $Z'$ bosons, as shown in Fig.~\eqref{ZZP}. As a result, even if there is no mass mixing at the tree level, we receive the oblique correction. In the second model, we introduced a doublet, which is charged under new symmetry, and hence it can mix $Z'$ boson with SM neutral bosons at tree level. This novel physics is characterized by oblique parameters in an effective lagrangian approach, as \cite{Strumia:2022qkt}
\begin{subequations}
\label{tedpole 2}
\begin{align}
& S= \frac{4\sin^{2}\theta_{w}}{\alpha_{Z}}\frac{2M_{W}^{2}g_{x}^{2}}{g^{2}g'^{2}M_{Z'}^{2}} \left[ Z_{e} -Z_{\phi} +Z_{L}   \right] \left[ g^{2}Z_{e}+g'^{2}(Z_{e}+2Z_{L}) \right],\\
& T=\frac{1}{\alpha_{Z}}\frac{4M_{W}^{2}g_{x}^{2}}{g^{2}M_{Z'}^{2}}\left[ Z_{e} - Z_{\phi} + Z_{L} \right]^{2},\\
& U = \frac{4\sin^{2}\theta_{w}}{\alpha_{Z}} \frac{4M_{W}^{2}g_{x}^{2}}{g^{2}M_{Z'}^{2}}\left[ Z_{e} -Z_{\phi} +Z_{L}   \right]\left[ Z_{e}+2Z_{L}   \right].
\end{align}
\end{subequations}
The $U(1)_{B-L}$ charges of the Lepton singlet, Lepton doublet are $Z_{e}$, $Z_{L}$ respectively, and $Z_{\phi}$ is the total $B-L$ charge of both the scalar doublets. $M_{Z'}$ is the mass of the new heavy boson, and $g_{x}$ is the new gauge coupling.
 \begin{figure}[ht]
   \centering
   \captionsetup{justification=raggedright}
  \includegraphics[width=0.4\textwidth]{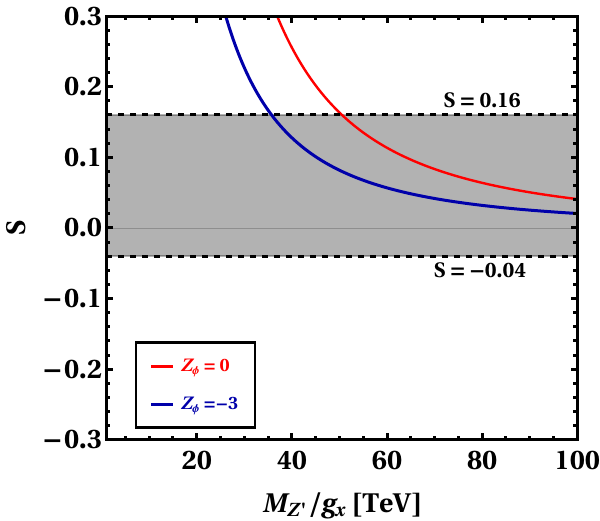}
  \includegraphics[width=0.4\textwidth]{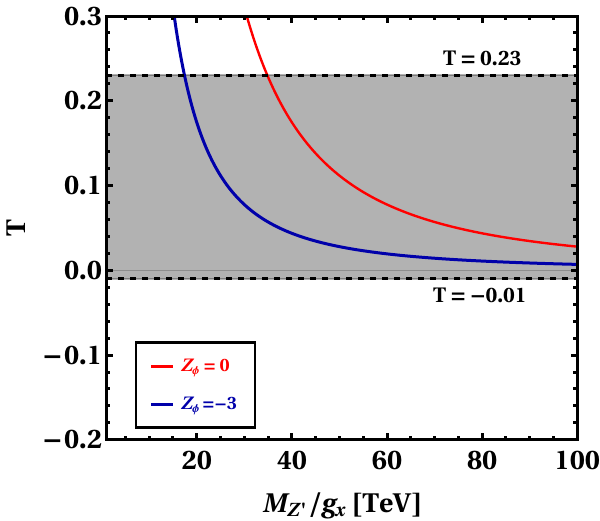}
  \includegraphics[width=0.4\textwidth]{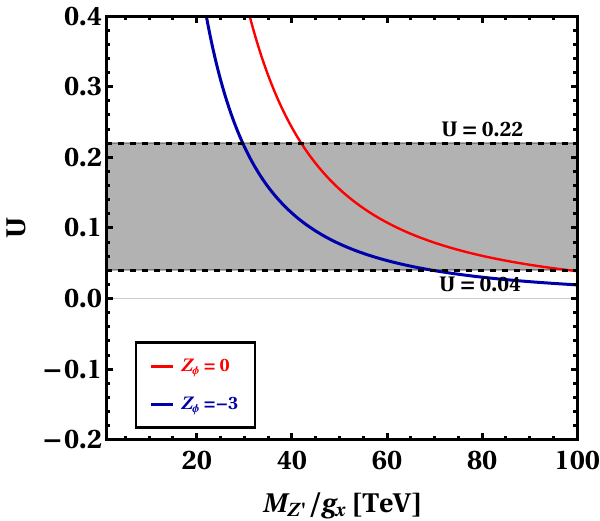}
  \caption{The $S$, $T$, and $U$ parameters versus the ratio of the mass of new boson and its gauge coupling. The red line represents the vector $B-L$ model, whereas the blue line represents the chiral $B-L$ model.}\label{STU}
\end{figure}
Taking the central values of the input parameters, $M_{W}^{CDF}=80.4335~\text{GeV}$~\cite{CDF:2022hxs},~ $\sin^{2}\theta_{w}=0.23143$~\cite{Lu:2022bgw}, and $\alpha_{Z}=1/127.935$~\cite{ParticleDataGroup:2020ssz, Babu:2022pdn} in Fig.~\ref{STU}, we plotted The $S, T,U$ parameters versus ratio of the mass of new boson and its gauge coupling for two different scenarios. The red line depicts the vector $B-L$ model, whereas the blue line depicts the chiral $B-L$ model.
The dotted lines correspond to the updated best-fit values of the $S,T,U$ parameters after the CDF results \cite{Lu:2022bgw}.
 \begin{figure}[ht]
   \centering
   \captionsetup{justification=raggedright}
  \includegraphics[width=0.49\textwidth]{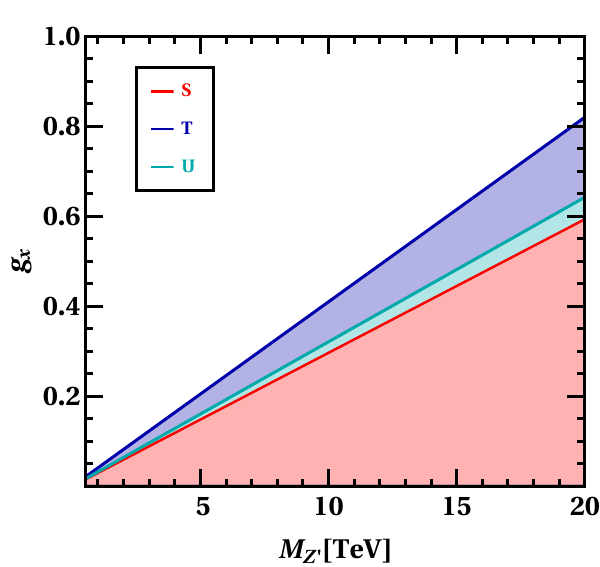}
  \includegraphics[width=0.49\textwidth]{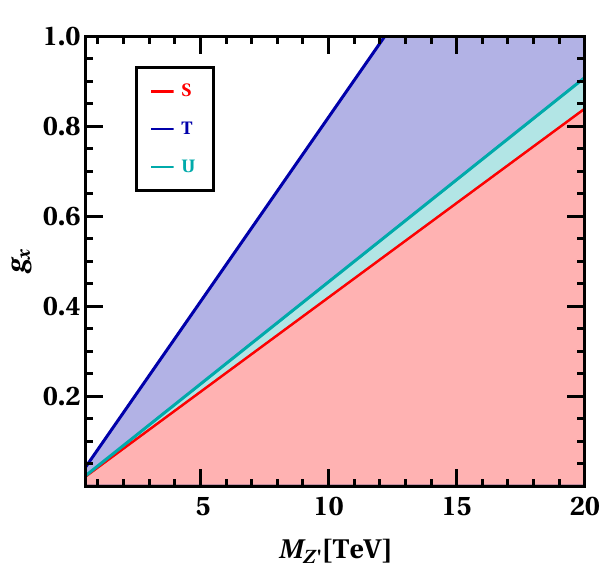}
  \caption{The new boson's gauge coupling versus its mass. The space permitted by the new $S,T, \text{and}~ U$ ($3\sigma$) is shown by a different coloured band. The graph on the left is for the vector $B-L$ ($Z_{\phi}=0$), while the graph on the right is for the chiral $B-L$ ($Z_{\phi}=-3$) model.}\label{STU2}
\end{figure}

In Fig.~\ref{STU2}, the parameter space that permits us to solve the $W$ anomaly is shown. The blue line represents the maximum permissible value for $T$, and the region between the blue line and the $M_{Z'}$ axis represents the allowable parameter space fulfilled by the best-fit $T$ value in the 3$\sigma$ range. The allowed region for $S$ and $U$ is shown by the colours red and cyan.
 \begin{figure}[ht]
   \centering
   \captionsetup{justification=raggedright}
  \includegraphics[width=0.49\textwidth]{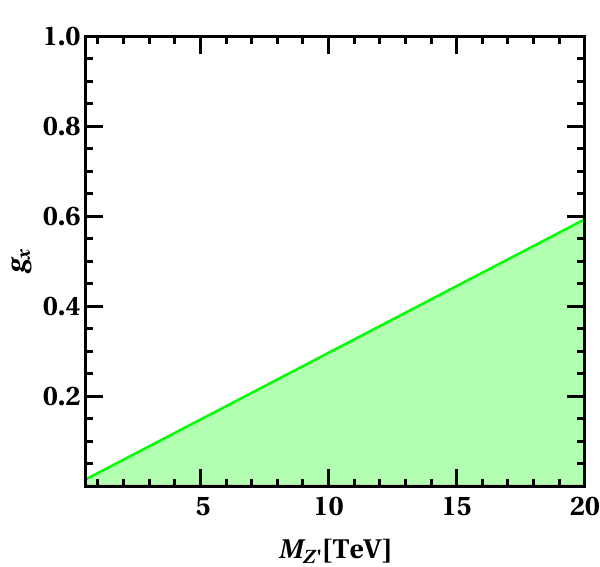}
  \includegraphics[width=0.49\textwidth]{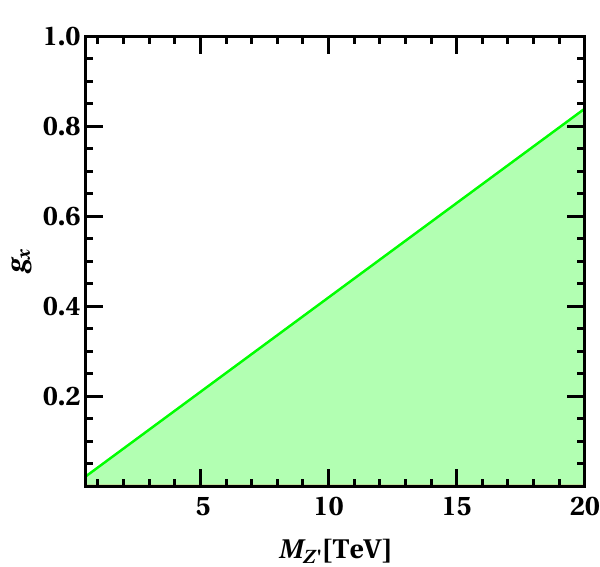}
  \caption{The green band represents the permitted parameter space that is consistent with CDF II, $W$ mass measurement. The graph on the left is for $Z_{\phi}=0$, while the graph on the right is for $Z_{\phi}=-3$. }\label{STU3}
\end{figure}

The allowed parameter space in Fig.~\ref{STU3} to satisfy CDF-II measurements is the overlap zone between the $S, \, T$ and $U$ bands. As previously stated, the left panel depicts the case with no mass mixing between SM neutral gauge bosons and the new $U(1)_{ B-L}$ neutral gauge boson, whereas the right panel shows the case with mass mixing as $\varphi$ has a charge of $-3$ under $B-L$ symmetry. Notice that the parameter space is improved when the scalar doublet mixes the new heavy boson with SM neutral bosons at tree level. As a result, we will now concentrate our efforts on the chiral $B-L$ model.

\begin{figure}[ht]
   \centering
   \captionsetup{justification=raggedright}
  \includegraphics[width=0.49\textwidth]{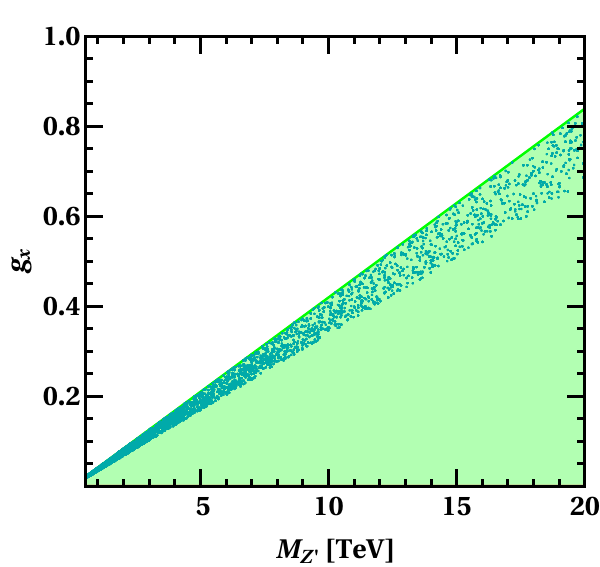}
  \caption{For the chiral $B-L$ model, we showed the space permitted by the new oblique parameters ($S, T, U$) after CDF-II results (3$\sigma$) in green colour. Dark Cyan points are consistent with CDF-II, $W$ mass measurement.}\label{STUW}
\end{figure}

Taking the scenario where $\varphi$ has a charge of $-3$ under $B-L$ symmetry, we showed the points~(dark cyan) that satisfy $W$ mass in Fig.~\ref{STUW}.
\section{Dark Matter Constraints}
\label{sec:dark-matter}
In this section, we collect the results of our analysis of DM phenomenology. As previously stated, in the second model outlined in the Sec.~\ref{sec:model}, $\chi_d$ is the scalar DM. It carries the $U(1)_{B-L}$ charge $1/2$, which forbids any term in the potential that results in the decay of $\chi_d$. We study the $\chi_d$ relic density and its direct detection prospects. Specifically, we determine the regions in the parameter space of the model where the DM constraints and $S$, $T$, $U$ consistent with CDF-II $W$ mass measurements can be satisfied. The SARAH-4.14.5 \cite{Staub:2015kfa,Goodsell:2017pdq} package is used to calculate all of the vertices and mass matrices, among other things. All the expressions are verified analytically and numerical calculations are performed by package SPheno-4.0.2 \cite{Porod:2003um,Porod:2011nf}. The relic abundance, on the other hand, is determined using micrOMEGAS-5.2.13~\cite{Belanger:2018ccd}.

\begin{figure}[ht]
   \centering
   \captionsetup{justification=raggedright}
  \includegraphics[width=0.85\textwidth]{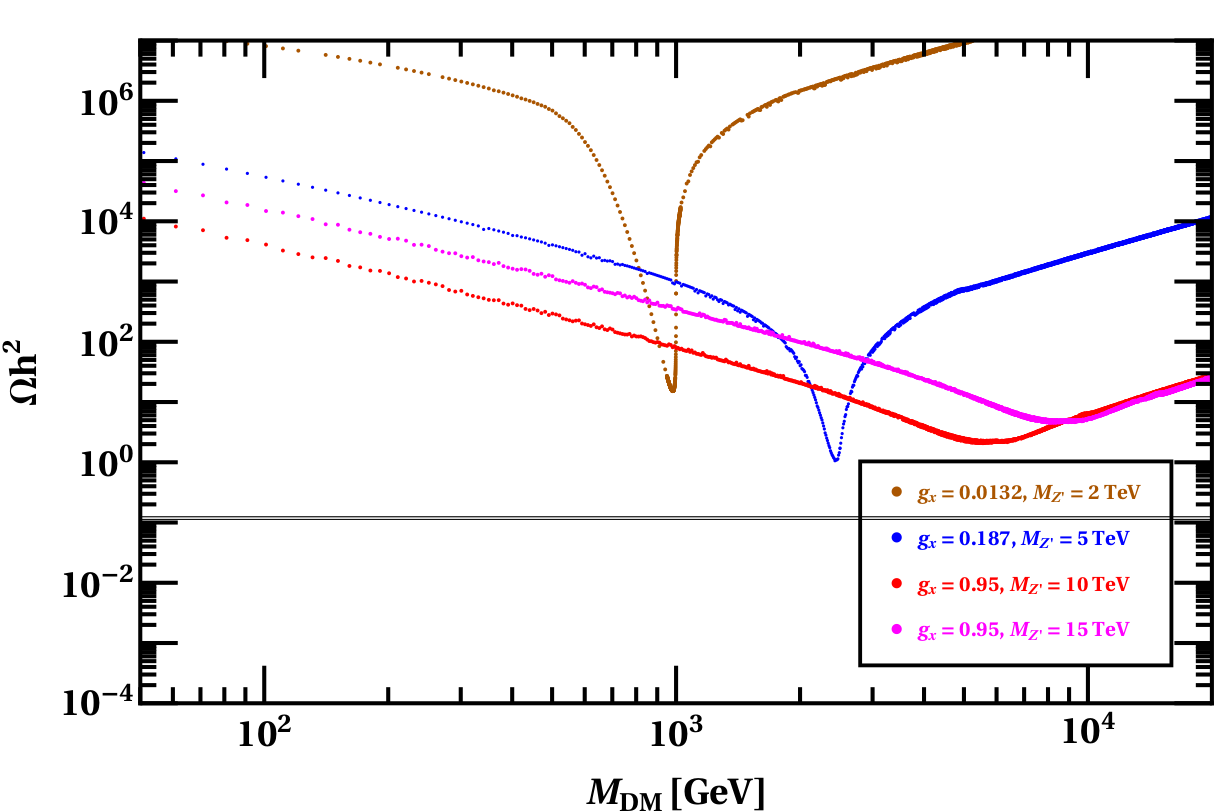}
  \caption{Relic density vs the mass of the DM for four different values of ($M_{Z'}, g_{x}$). We considered that only gauge interactions contribute to the relic density. The narrow horizontal band is the $3\sigma$ range for cold DM derived from the Planck satellite data~\cite{Planck:2018vyg}.} \label{OnlyZBM}
\end{figure}

There are several DM annihilation channels present in this model, which are shown in Appendix.~\ref{appn}. They involve annihilation to quarks, leptons, neutrinos, gauge bosons~($Z$, $Z'$), neutral scalars~($H_{i}, H^0$) and charged scalar~($H^{\pm}$). Altogether, they determine the relic abundance of $\chi_d$. Note that as $\chi_d$ is charged under $B-L$, the DM $\chi_d$ has both the gauge and scalar interactions. The gauge interactions allow the annihilation of the DM particle into fermions mediated by the gauge boson, $\chi_d\chi_d^{*}\to Z'\to f\bar{f}$. One should also consider the direct annihilation into two gauge bosons, $\chi_d \chi_d^{*}\to Z' Z'$, when kinematically accessible. Hence, in a pure gauge interaction case, there are very few parameters~($M_{\rm DM}, g_x, M_{Z'}$) which play a role in determining DM phenomenology. Due to the strong experimental constraints on $M_{Z'}/g_x$, the annihilation into $Z' Z'$ is suppressed. Hence, only annihilation to fermions turns out to be relevant. Due to the structure of the gauge coupling, the annihilation channel $\chi_d\chi_d^{*}\to Z'\to f\bar{f}$ is velocity suppressed~($\propto v^2$). This is why the relic density tends to be much higher than the observed value except in a narrow region close to the resonance~($M_{\rm DM}\sim M_{Z'}/2$)~\cite{Rodejohann:2015lca}. And even at the resonance, the relic density can be too large to be in agreement with the data, as illustrated in Fig. \ref{OnlyZBM}.
 In this Figure, we considered that only gauge interactions contribute to relic density and plotted it as a function of DM mass. We considered four benchmark values of ($M_{Z'},g_{x}$). We chose the maximum gauge coupling as the benchmark coupling value that satisfies the collider bounds, as shown in Fig.~\ref{fig:B-LConstraint}, for DM masses of 2 TeV and 5 TeV. For the other two cases, we used a gauge coupling of 0.95. It is clear from Fig.~\ref{OnlyZBM} that the higher the $Z'$ mass, the more difficult it is to get low values of the relic density, even for very high gauge coupling.
 \begin{figure}[ht]
   \centering
   \captionsetup{justification=raggedright}
  \includegraphics[width=0.9\textwidth]{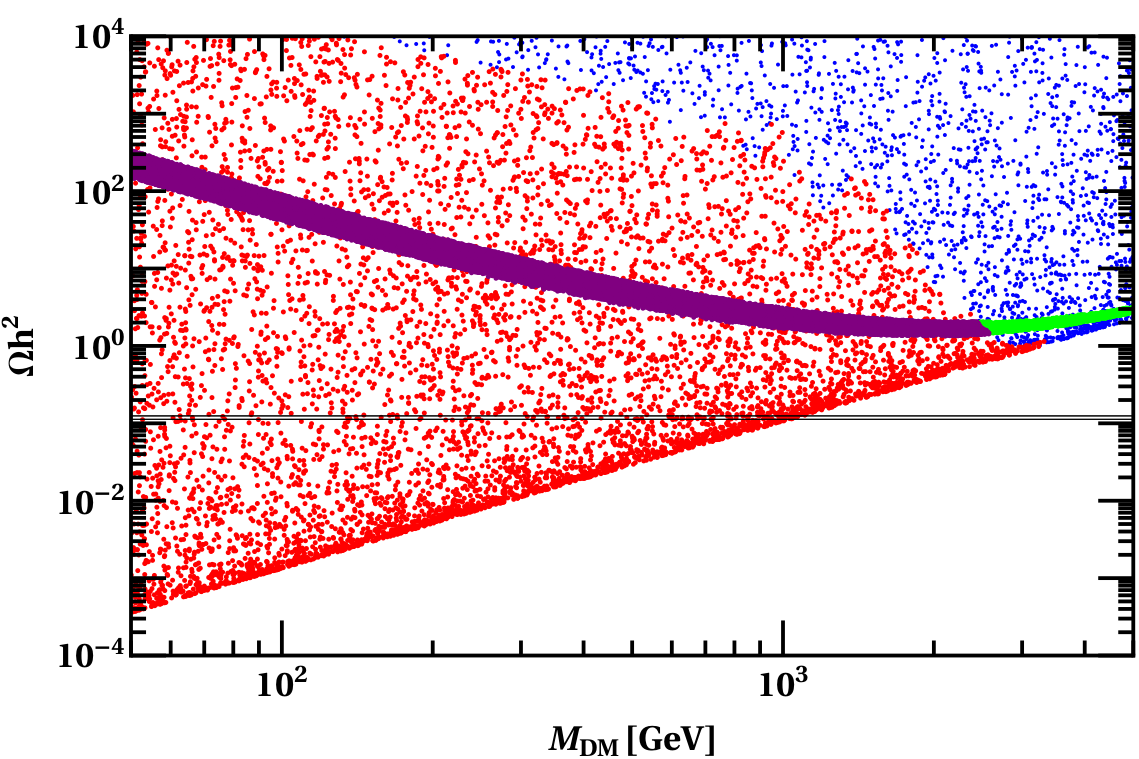}
  \caption{Relic density vs the mass of the DM. The narrow horizontal band is the $3\sigma$ range for cold DM derived from the Planck satellite data~\cite{Planck:2018vyg}. We considered that only gauge interactions contribute to the relic density. We varied gauge coupling from $10^{-4}$ to 1, and $Z'$ mass in the range of 0.1 TeV to 10 TeV. Purple-Green band satisfy the STU and CDF-II W mass at $3\sigma$. ATLAS and CMS dilepton searches rule out the red and purple points but allow the green and blue points. }\label{OnlyZScatter}
\end{figure}
\par Fig.~\ref{OnlyZBM} indicates that the only hope to reproduce the known DM relic abundance is to be near the $Z'$ resonance, i.e. $M_{\rm DM}\sim M_{Z'}/2$. Assuming this is the case we did a full numerical scan of the whole parameter space. We show this in Fig.~\ref{OnlyZScatter} along with the constraints on the $M_{Z'}/g_{x}$ from the collider and the W mass anomaly. We varied gauge coupling from $10^{-4}$ to 1, and $Z'$ mass in the range of 0.1 TeV to 10 TeV. Purple-Green band satisfy the S, T, U and CDF-II W mass at $3\sigma$, the range of coupling for this band in $M_{Z'}-g_{x}$ plane is shown in Fig.~\ref{STUW}. ATLAS and CMS dilepton searches rule out the red and purple points but allow the green and blue points. It is clear from Fig. \ref{OnlyZScatter} that the requirements to have a correct relic is to be near resonance and relatively large coupling $g_x$ and low $M_{Z'}$, which in view of Fig.~\ref{fig:B-LConstraint} is ruled out.
\par Hence, due to strong constraints on $M_{Z'}/g_{x}$ from the ATLAS and CMS dilepton searches, it is not possible to satisfy the relic density purely through the gauge sector. Additionally, the $M_{Z'}/g_{x}$ range needed to satisfy the CDF-II, W mass causes the relic density to be excessively large, even near the resonance.

\begin{table}[ht]
\centering
\captionsetup{justification=centering}
 \begin{tabular}{|c|c|}
 \hline \hline \centering
 ~~~~~~~~Parameter~~~~~~~ & ~~~~~~~Range~~~~~~~~   \\ 
 \hline 
 \hline
 $m_{\chi_{d}}$ & ~~~~~$[~0,~10^{4}~]$~GeV \\ 
 \hline 
 $\lambda_{\Phi\chi_{d}}$ & $[~10^{-6},~1 ~]$ \\  
  \hline 
 $\lambda_{\varphi \chi_{d}}$ & $[~10^{-6},~1~ ]$ \\  
 \hline 
 $\lambda_{\sigma \chi_{d}}$ & $[~10^{-6},~1~ ]$ \\  
 \hline
  $\mu$ & ~~~~~~~$[~10^{-6},~1~ ]$~GeV \\  
 \hline 
 \end{tabular}  
\caption{Ranges of variation of the input parameters used in our numerical scan.}
\label{parametertable}
\end{table}
Besides the gauge interactions, the DM $\chi_d$ also has scalar interactions; see Eq.~\ref{eq:chiral-potential}. Scalar interactions between the DM and the SM scalar give rise
to the well-known Higgs-portal scenario. But there are some differences between this simplistic Higgs-portal scenario and our $B-L$ case. First of all, the DM field is necessarily complex as it is charged under $U(1)_{B-L}$ – rather than real. In addition to this, there will be many additional annihilation channels due to the presence of additional scalars, both neutral and charged, such as $H_{2,3}$, $H^{0}$ and $H^{\pm}$. 
\par In the following, instead of separately studying gauge interaction and scalar interaction, we will focus on the general case. To do this we first need to identify the parameters which are relevant for DM analysis. Even though the model introduces new free parameters, not all of them are important to DM analysis. For example, the self-quartic couplings and some mixed quartic couplings such as $\lambda_{\Phi\sigma},\lambda_{\varphi\sigma}, \lambda_{\Phi\varphi_{1,2}}$ does not play any role in DM phenomenology. Hence, we choose to fix these parameters. The remaining free parameters relevant for DM analysis can be chosen as:
\begin{align}
m_{\chi_d}, \lambda_{\Phi \chi_d}, \lambda_{\varphi\chi_d}, \lambda_{\sigma\chi_d}, g_x\,\,  \text{and} \,\, M_{Z'}.
\end{align} 
We will look at how the DM phenomenology of this model is affected by the above-mentioned parameter. To carry out the numerical scan, we varied these parameters as listed in Table.~\ref{parametertable}. We varied them on the logarithmic scale. The gauge coupling $g_{x}$ and mass $M_{Z'}$ are varied according to the allowable parameter space coming from $S$, $T$, $U$ restriction consistent with CDF-II $W$ mass measurements (Dark cyan points ), as illustrated in Fig.~\ref{STUW}. 
\begin{figure}[ht]
  \includegraphics[width=0.8\textwidth]{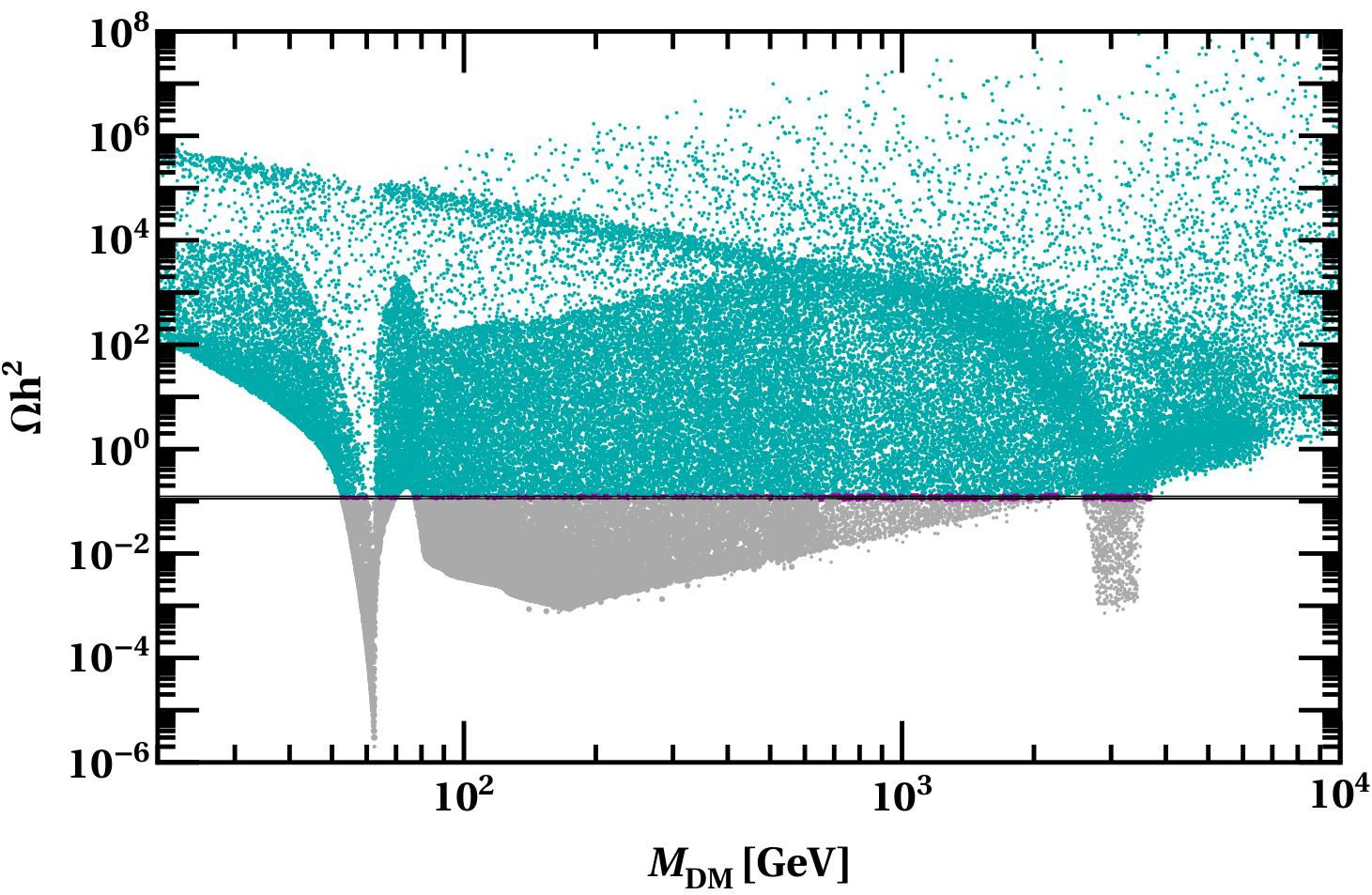}
  \caption{Relic density vs the mass of the DM. The darker cyan and gray dots show over- and under-abundance relic density points, respectively, while the magenta points reflect the 3 $\sigma$ range for cold DM obtained from Planck satellite data~\cite{Planck:2018vyg}.}
  \label{fig:Relic}
\end{figure}
\par In Fig.~\ref{fig:Relic}, we show the relic density as a function of the mass of the scalar DM $\chi_d$. The darker cyan and gray points represent the over and under-abundance relic density regions, respectively, whereas the magenta points in the narrow band fall in the $3\sigma$ range for cold DM derived from the Planck satellite data~\cite{Planck:2018vyg}:
\begin{equation}
0.1126 \leq \Omega h^2 \leq 0.1246.
\end{equation}
Various features of Fig.~\ref{fig:Relic} can be understood from different DM annihilation channels shown in Appendix.~\ref{appn}. Annihilation in the low mass region of DM is dominated via the exchange of SM Higgs~($H_{1}=h$) to SM fermionic final states. As DM approaches half of the Higgs mass ($~M_{\rm DM}\approx m_h/2$), $h$ becomes on-shell, and these annihilation channels become very efficient. Notice that there is no dip at $M_{\rm DM} \approx M_{Z}/2$ because the mixing between $Z$ and $Z'$ is not strong enough for annihilation through $Z$ exchange to be effective.  For $M_{\rm DM} \geq 80 ~\text{GeV}$, annihilation of DM to $Z$ and $W$ final states comes into the picture  ($\chi_{d} \chi_{d}^{*} \rightarrow Z Z,~ W^{+} W^{-} $) and hence we get another dip in that region. We see a dip in relic density in the DM mass range  2.5 TeV-4 TeV. This is due to the fact that the combination of $g_{x}$ and $M_{Z'}$ required to get correct oblique parameters forces the vev of singlet scalar ($v_{\sigma}$) to be around 8 TeV-10 TeV. This high vev pushes the mass of CP-even scalars $H_{2}$ and $H_{3}$ to be in the range of 5-8 TeV and we get a dip when DM mass is roughly half of this range due to the $H_{2,3}$ mediated s-channel annihilation to SM final states. A sub-dominant role is played by annihilation into $H_i H_j$, $H^0H^0$ and $ZZ$, $Z' Z'$ via the direct 4-point vertices $H_{i}H_{j}\chi_d\chi_d^*$, $H^0H^0\chi_d\chi_d^*$ and $ZZ\chi_d\chi_d^*$ or $ZZ'\chi_d\chi_d^*$, respectively. Also, there could be an additional contribution from $\chi_d$ exchange in the t-channel. As the annihilation cross section is inversely proportional to the mass of the DM, at a very high value of DM mass the relic density increases.

\par \underline{\bf Direct Detection:}~ Let us now study the direct detection prospects of our DM candidates $\chi_d$. A large number of experiments are being conducted to demonstrate the particle nature of DM through direct detection. Various direct detection experiments, XENON1T \cite{XENON:2018voc}, LZ \cite{LZ:2022lsv}, XENONnT\cite{XENON:2022ltv}, LUX\cite{LUX:2013afz,LUX:2016ggv}, PandaX-II\cite{PandaX-II:2020oim}, impose constraints. These experiments are designed to measure the tiny recoil in the detector target nuclei produced by the elastic collisions between DM and target nuclei.\\
The effective lagrangian for nucleon-DM interactions is expressed as
\begin{equation}
\label{efflag}
\mathscr{L}_{eff}= a_{N}\overline{N}N \chi_{d}^{2},
\end{equation}
Where $a_{N}$ is the effective nucleon-DM coupling. The spin-independent scattering cross section via the Higgs$(H_{1,2,3})$ interaction is given by
\begin{equation}
\label{Higgscrossscetion}
\sigma_{N-\chi_{d}}^{SI} = \frac{\mu^{2}M_{N}^{2}f_{N}^{2}}{4 \pi M_{DM}^{2} v^{2}} \left[ \frac{\lambda_{H_{1} \chi_{d}^{2}}}{M_{H_{1}}^{2}}(\mathcal{O}_{R})_{11} +  \frac{\lambda_{H_{2} \chi_{d}^{2}}}{M_{H_{2}}^{2}}(\mathcal{O}_{R})_{21}+ \frac{\lambda_{H_{3} \chi_{d}^{2}}}{M_{H_{3}}^{2}}(\mathcal{O}_{R})_{31}  \right],
\end{equation}
Where $(\mathcal{O}_{R})_{ij}$ is the elements of the mass matrix defined in Eq.~\eqref{rot1}, $f_{N}$ is the form factor, which depends on the hadronic matrix elements and $\mu=\frac{M_{N}M_{DM}}{M_{N}+M_{DM}}$ is the reduced mass for nucleon-DM system. The trilinear couplings are given as
\begin{equation}
\lambda_{H_{i} \chi_{d}^{2}} = 2 \left[v_{\Phi} \lambda_{\Phi \chi_{d}}  (\mathcal{O}_{R})_{i1}      +  v_{\varphi}  \lambda_{\varphi \chi_{d}}  (\mathcal{O}_{R})_{i2}+         v_{\sigma}  \lambda_{\sigma \chi_{d}}  (\mathcal{O}_{R})_{i3} \right] .
\end{equation}
Eq.~\eqref{Higgscrossscetion} is an extension of the expression corresponding to the scalar DM case \cite{Cline:2013gha}. The cross-section per nucleon for DM-nuclei interaction through $Z'$ is given as \cite{Ma:2015mjd}
\begin{equation}
\label{crosssection}
\sigma_{0} = \frac{1}{\pi} \left( \frac{M_{DM} M_{n}}{M_{DM}+M_{n}} \right)^{2} \left( \frac{g_{x}}{M_{Z'}}  \right)^{4},
\end{equation}
Where $A$ is the number of nucleons in the target, we have set it to $131$ for Xenon. The nucleon mass is $M_{n}=0.938919$~GeV.
\begin{figure}[ht]
   \centering
   \captionsetup{justification=raggedright}
  \includegraphics[width=0.8\textwidth]{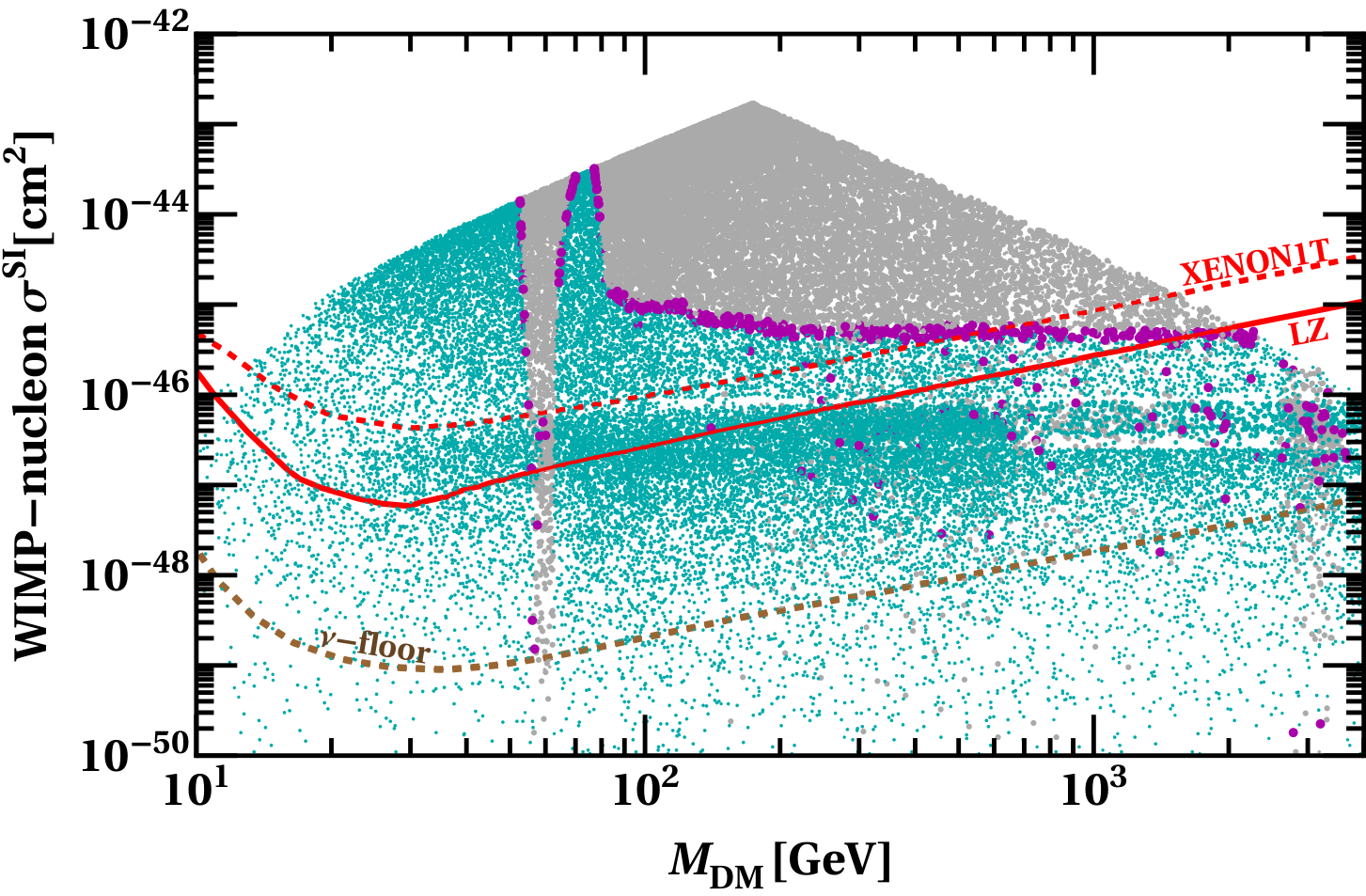}
  \caption{WIMP-nucleon spin-independent cross section for the scalar DM $\chi_{d}$. The colour code has the same meaning as in Fig.~\ref{fig:Relic}. The solid red line denotes the latest upper bound from the LZ \cite{LZ:2022lsv} collaboration, the dashed red line corresponds to XENON1T \cite{XENON:2018voc} limit, and the dashed brown line corresponds to the “neutrino floor” lower limit \cite{Billard:2013qya, Billard:2021uyg}. }\label{DD}
\end{figure}
\par In Fig.~\ref{DD}, we imposed the direct detection constraints on our scalar DM $\chi_{d}$. We performed the numerical scan with the micrOMEGAS-5.2.13 and varied the parameters as shown in Table.~\ref{parametertable}. The colour code has the same meaning as in Fig.~\ref{fig:Relic}. The LZ and XENON1T experiment imposes the most stringent constraints. As a result, we plotted the most recent upper bound from both LZ and XENON1T collaboration~\cite{XENON:2018voc}, as shown by the solid and dashed red lines, respectively. The brown line represents the lower limit, which corresponds to the ``neutrino floor" from the coherent elastic neutrino scattering~\cite{Billard:2013qya, Billard:2021uyg}.
\par
\begin{figure}[ht]
   \centering
   \captionsetup{justification=raggedright}
  \includegraphics[width=0.49\textwidth]{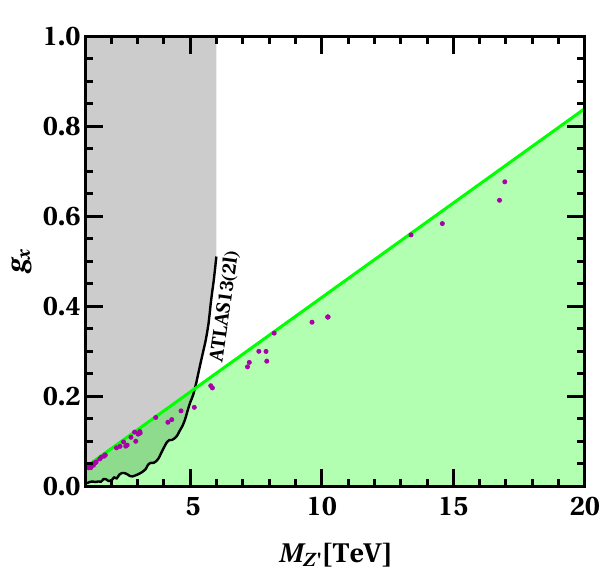}
  \caption{We showed the space permitted by the new oblique parameters ($S, T, U$) after CDF-II results in green colour. The gray region is ruled out by the most stringent collider constraints from ATLAS. The magenta points satisfy $M_{W}^{CDF}$, relic density and limits from direct detection experiments simultaneously.}\label{ALLLZ}
\end{figure}

 In Fig.~\ref{ALLLZ}, we demonstrated the parameter space that is compatible with all of the aforementioned requirements. The green bands show the permitted values for oblique parameters $S, T$ and $U$. The region with the gray shading is the ATLAS's most stringent collider constraint. The magenta points fulfil $M_{W}^{CDF}$, relic density and limitations from direct detection experiments all at the same time.

\par It is worth noting that there is a significant parameter space in the DM high mass region, which is consistent with recent measurements of W boson mass, relic abundance, collider constraints, and the direct detection experiments.    

\section{Dark Matter Constraints in View of Recent CMS and ATLAS Results}\label{Sec.ATLAS}
Recently, the CMS Collaboration reported the first $W$-boson mass measurement, $M_{W}^{\mathrm{CMS}} = 80.3602 \pm 0.0099~\text{GeV}$~\cite{CMS:2024lrd}, using the data sample collected in 2016 from proton--proton collisions at $\sqrt{s} = 13~\text{TeV}$. In addition, the ATLAS Collaboration presented an improved measurement, $M_{W}^{\mathrm{ATLAS}} = 80.3665 \pm 0.0159~\text{GeV}$ \cite{ATLAS:2024erm}, based on the data recorded in 2011 at $\sqrt{s} = 7~\text{TeV}$.  Both the ATLAS and CMS measurements are consistent with the SM prediction, with the CMS result exhibiting a smaller uncertainty than that of ATLAS.

To assess the impact of these measurements, we have re-analysed our data using the global electroweak fits reported in Ref.~\cite{Lu:2022bgw}. In this work, the authors present electroweak fits performed separately with the CDF~II result and the PDG~2021 value, thereby enabling a direct comparative analysis. The oblique parameters  at $1 \sigma$ consistent with PDG measurements are given as
\begin{equation}
S = 0.06 \pm 0.10,~ T = 0.11 \pm 0.12,~U = -0.02 \pm 0.09, 
\end{equation}
with correlation $\rho_{ST} =0.9,~\rho_{SU} = -0.57$, and $\rho_{TU} = -0.82$.
We repeat our analysis under the PDG constraints on the oblique parameters, ensuring consistency with the $W$-boson mass reported by CMS, $M_{W}^{\mathrm{CMS}} = 80.3602 \pm 0.0099~\text{GeV}$~\cite{CMS:2024lrd}, as this measurement has a smaller uncertainty than the corresponding ATLAS result.
 \begin{figure}[ht]
   \centering
   \captionsetup{justification=raggedright}
  \includegraphics[width=0.49\textwidth]{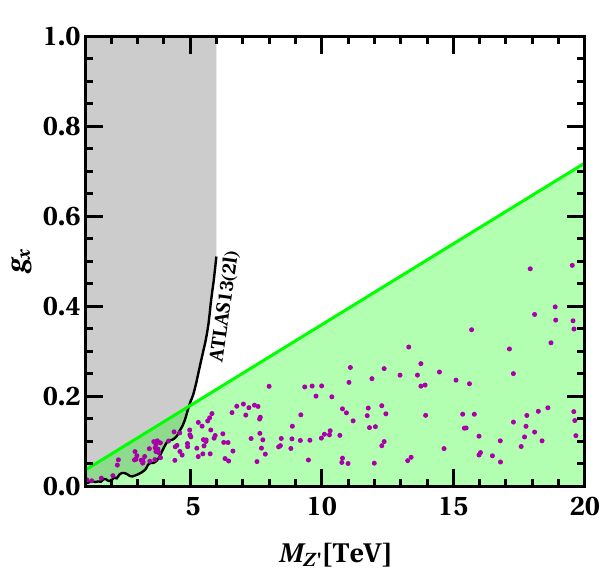}
  \caption{Region permitted by the PDG oblique parameters at $3 \sigma$ in green. The gray-shaded area is excluded by the collider constraints from ATLAS. The magenta points simultaneously satisfy $M_{W}^{\mathrm{CMS}}$, the relic density, and the limits from direct detection experiments.
}\label{CMS}
\end{figure}
In Fig.~\ref{CMS}, we present the parameter space compatible with the PDG oblique parameters at $3 \sigma$, the CMS $W$-boson mass at $3 \sigma$, collider constraints, relic density, and direct detection experiments. The green bands show the permitted values for the oblique parameters. The gray-shaded region is ruled out by ATLAS's most stringent collider constraints. The magenta point satisfies the CMS $W$-boson mass measurement and is also consistent with the measured relic density and the constraints from direct detection experiments.
Comparing Fig.~\ref{ALLLZ} and Fig.~\ref{CMS}, it is evident that the allowed parameter space corresponding to the CMS $W$-boson mass measurement is significantly broader than that for the CDF II result. Finally, it should be noted that the CDF-II measurement itself is not consistent with the ATLAS and CMS measurements. As a result, there is no common parameter space in the model that is simultaneously consistent with both results. This, however, is not a shortcoming of the model but rather is an artifact of the incompatibility of the two measurements.

\section{Conclusion}
The $U(1)_{B-L}$ gauged extension of the SM is very simple in its nature. Its minimal version just needs three right-handed neutrinos to cancel gauge anomalies. It naturally explains the small neutrino masses through a seesaw mechanism. Despite their simplicity, these types of models can explain the recent CDF-II measurement of the $W$ boson mass, which reveals considerable disagreement with the SM predictions. In this work, we argued that the CDF~II $W$-boson anomaly can be viewed as originating from a deviation in the $Z$-boson mass, which, through the oblique parameters, manifests as the observed shift in the $W$-boson mass. New-physics effects can reconcile the $Z$-boson mass with its measured value, effectively transferring the anomaly to the $W$-boson sector. We show that the new neutral boson associated with the new $U(1)_{B-L}$ symmetry can provide the loop corrections to gauge boson two-point functions that are compatible with the most recently revised oblique parameter values as a consequence of the CDF-II results.

We investigated and found that kinetic mixing alone can not explain the $W$ anomaly at the tree level. In addition, we investigated the two models with and without mass mixing between $Z$ and $Z'$. We analyzed these models by simultaneously demanding that :
\begin{itemize}
    \item Model parameters satisfy S, T, and U parameters consistent with electroweak global fit considering CDF II $W$ boson mass measurements.
    \item Model is consistent with stability and perturbativity constraints.
    \item  Constraints on $Z'$ boson from LEP and LHC are satisfied.
\end{itemize}
We find that the parameter space corresponding to the chiral $B-L$ model is broader than that of the vector $B-L$ model; therefore, we focus on the chiral $B-L$ scenario to study the dark matter phenomenology. We impose constraints from dark matter direct detection experiments and relic abundance measurements, and demonstrate that the chiral $U(1)_{B-L}$ model can simultaneously satisfy these dark matter constraints and account for the $W$-boson mass anomaly.

\begin{acknowledgments}
  Work of S.M. has been supported by KIAS Individual Grants (PG086002) at Korea Institute for Advanced Study. The work of R.S.  is supported by the Government of India, SERB Startup Grant SRG/2020/002303. The work of H.P. is supported by the Prime Minister Research Fellowship (ID: 0401969).
\end{acknowledgments}
\appendix

\section{Annihilation channels for scalar DM $\chi_d$}
\label{appn}
\begin{figure}[ht]
   \centering
   \captionsetup{justification=raggedright}
  \includegraphics[width=0.49\textwidth]{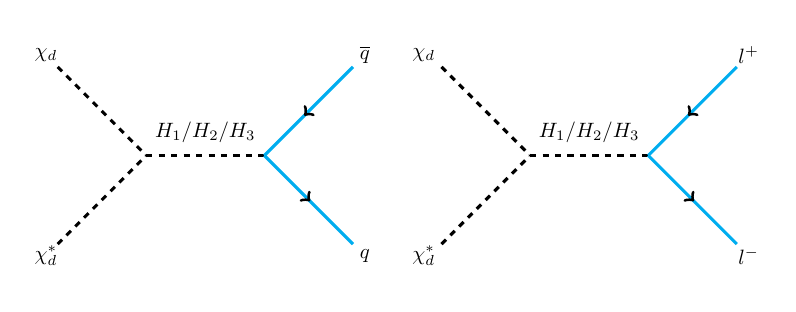}
   \includegraphics[width=0.49\textwidth]{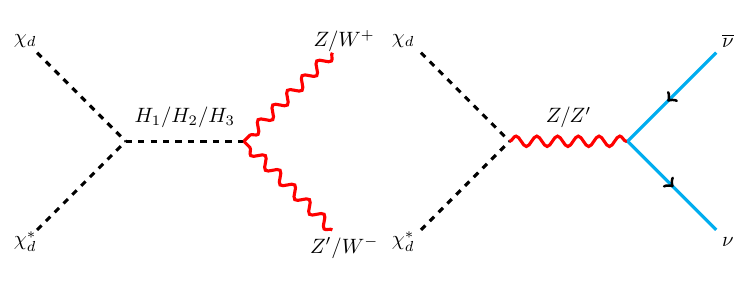}
    \includegraphics[width=0.49\textwidth]{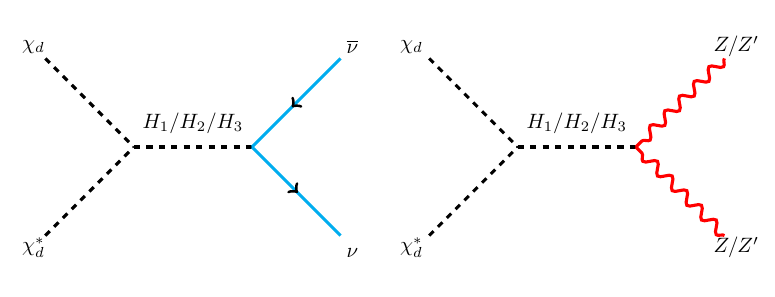}
     \includegraphics[width=0.49\textwidth]{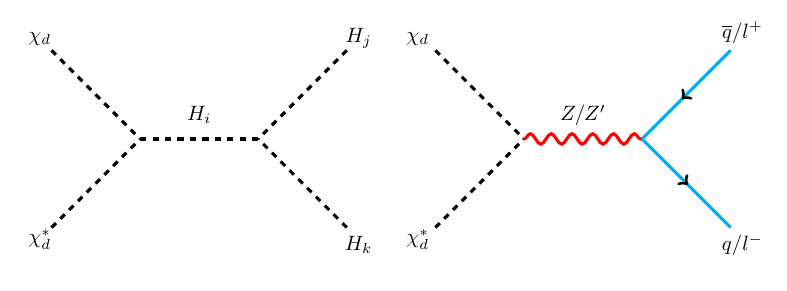}
      \includegraphics[width=0.49\textwidth]{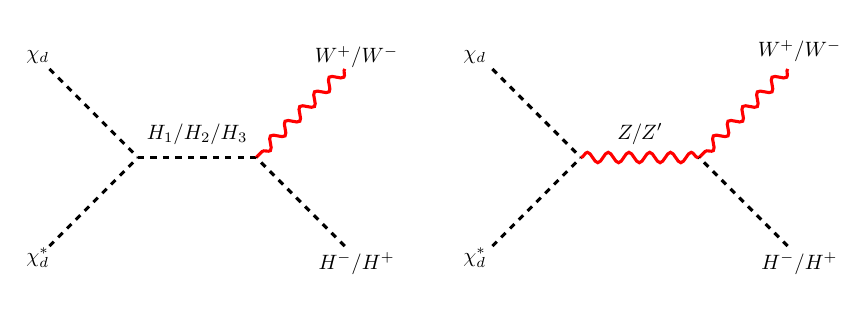}
       \includegraphics[width=0.49\textwidth]{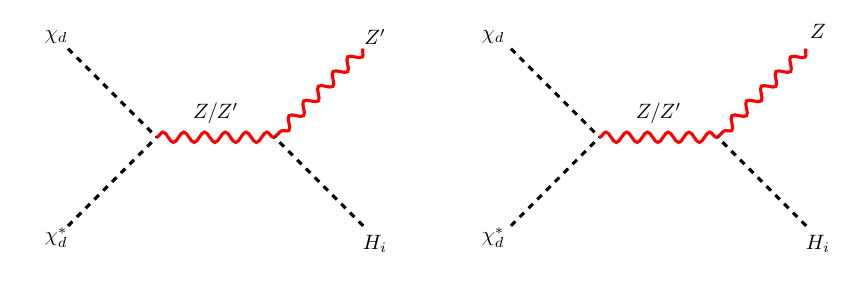}
                \includegraphics[width=0.8\textwidth]{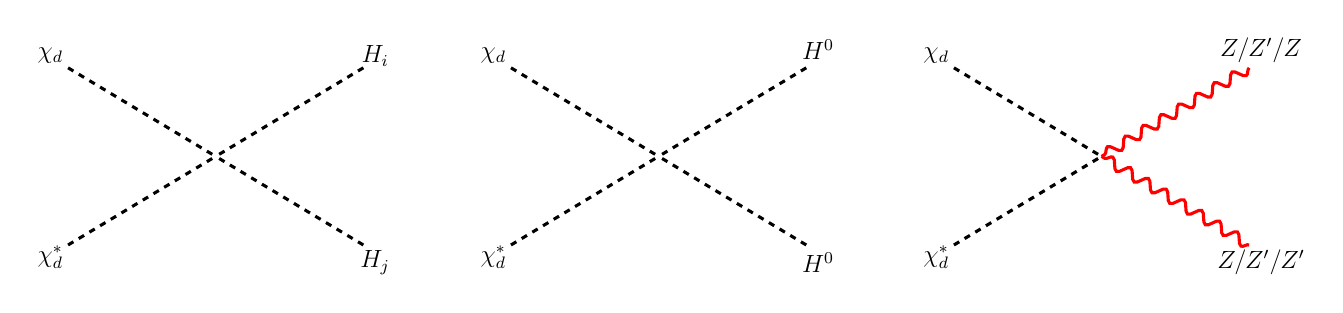}  
        \includegraphics[width=0.54\textwidth]{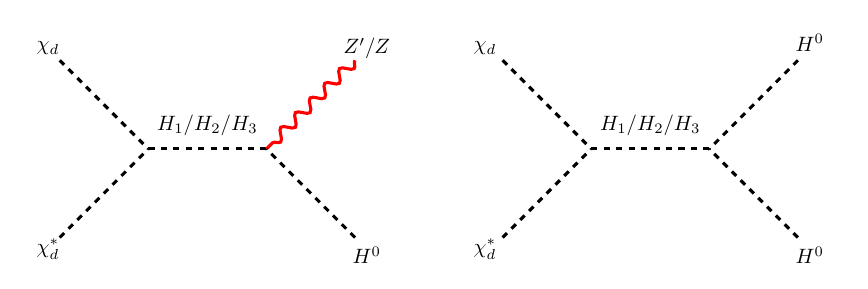}
        \includegraphics[width=0.26\textwidth]{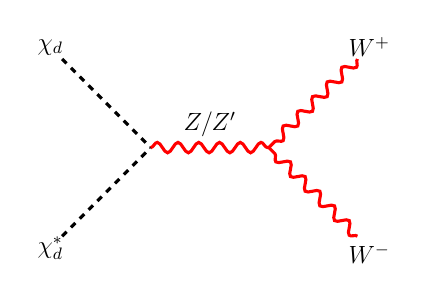}    
             \includegraphics[width=0.45\textwidth]{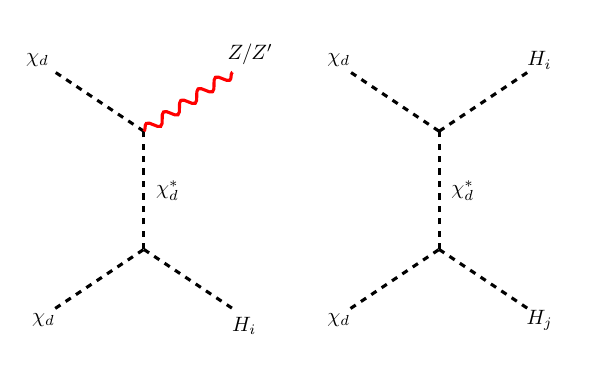}
       \includegraphics[width=0.25\textwidth]{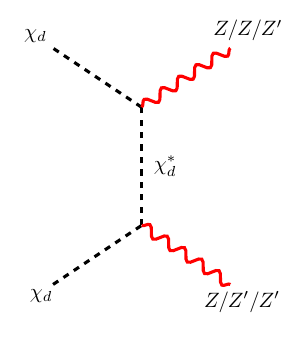}
  \caption{Feynman diagrams that contribute to the relic
density of the scalar DM $\chi_{d}$.}\label{DMfdia}
\label{fig:ann}
\end{figure}

In the chiral $B-L$ model the relic abundance of the DM candidate $\chi_d$ is determined by the annihilation diagrams shown in Fig.~\ref{fig:ann}.
\newpage
\bibliographystyle{utphys}
\bibliography{bibliography}
\end{document}